\documentclass[]{aa} 

\usepackage{listings}

\usepackage{xcolor}
\usepackage{graphicx}
\usepackage{txfonts}
\usepackage{amsmath}
\usepackage{yhmath}
\def\f{{Fig.~}}

\def\ff{{Figure~}}
\def\s{{Sect.~}}
\def\app{{Appendix~}}

\def\e{{Eq.~}}

\def\t{{Table~}}

\def\kpc{{\rm kpc}}
\def\kms{{\rm km\,s^{-1}}}
\def\kmskpc{{\rm km\,s^{-1}\,kpc^{-1}}}

\def\Gaia{{\it Gaia }}

\def\zc{$Z_{\rm{c}}$}
\def\zcm{Z_{\rm{c}}}

\def\vzc{$V_{Z\,{\rm{c}}}$}

\begin{document}

   \title{The phase spiral in Gaia DR3}


   \author{T. Antoja
          \inst{1,2,3}
          \and P. Ramos\inst{1,2,3,4} \and B. García-Conde\inst{5} \and M. Bernet\inst{1,2,3} \and C. F. P. Laporte\inst{1,2,3} \and D. Katz\inst{6}
          }
  \institute{Departament de Física Qu\`antica i Astrof\'isica (FQA), Universitat de Barcelona (UB),  c. Mart\'i i Franqu\`es, 1, 08028 Barcelona, Spain
           \email{tantoja@fqa.ub.edu}
    \and{Institut de Ci\`encies del Cosmos (ICCUB), Universitat de Barcelona (UB), c. Mart\'i i Franqu\`es, 1, 08028 Barcelona, Spain}
    \and{Institut d'Estudis Espacials de Catalunya (IEEC), c. Gran Capit\`a, 2-4, 08034 Barcelona, Spain} 
         \and{(Research Affiliate) National Astronomical Observatory of Japan, Mitaka-shi, Tokyo 181-8588, Japan}
         \and{Departamento de Física de la Tierra y Astrofísica, Fac. CC. Físicas, Universidad Complutense de Madrid, Plaza de las Ciencias, 1, Madrid, E-28040, Spain}
         \and{GEPI, Observatoire de Paris, Universit\'e PSL, CNRS, 5 Place Jules Janssen, 92190 Meudon, France}\\
             }

   \date{Received 21 November 2022 / Accepted 7 March 2023}

 
  \abstract
   {}
   {We aim to study the phase spiral in the Milky Way (MW) disc with data from the third data release of \Gaia (DR3) and use it as an inference tool to decipher the late-time evolution of the Galaxy.}
   {We used an edge-detection algorithm to find the border of the phase spiral, allowing us to robustly quantify its shape at different positions and for different selections. We calculated the time of onset of the phase-mixing by determining the different turns of the phase spiral and using the vertical frequencies from commonly used models of the gravitational potential of the MW.}
   {We find that the phase spiral extends down to $-1.2\,\kpc$ in height below the plane (about 3 to 5 scale heights of the thin disc) and beyond $\pm 50\,\kms$ in $V_Z$. We see a secondary branch mostly at positive vertical velocities when coloured by azimuthal velocity and in the counts projection. We also find complex variations of the phase spirals with angular momentum and azimuth. All these findings are possible evidence of multiple perturbations (from different times or from different perturbers) and/or of the complexity of the  phase-mixing process. We detect the phase spiral from 6 to 11 kpc from the Galactic centre and find signatures of vertical asymmetries 1-2 kpc beyond this range. We measure small but clear variations with azimuth. When we determine the phase mixing times from the phase spiral at different angular momenta and using the different spiral turns (at different $Z$), we obtain inconsistent times with systematic differences (times increasing with $|L_Z|$ and with $|Z|$). Our determinations are mostly in the range of [0.3-0.9] Gyr, with an average of 0.5 Gyr. The inconsistencies do not change when using different commonly used potential models for the MW, different stellar distances, or frequencies for different kinetic temperatures; they could stem from the inconsistency of the assumed gravitational potentials with the true MW, and from oversimplification of the modelling, in particular where self-gravity is neglected or where multiple perturbations and/or interference with other
processes are not considered.}
   {The wealth of information provided by the new \Gaia DR3 data should encourage us to make progress in crucial modelling aspects of the disc dynamics, such as non-equilibrium, self-gravity, propagation of different types of bending waves, and interactions between different mechanisms.  Such advancements could finally enable us to establish the origin of the phase spiral and its relation to the Sagittarius dwarf galaxy.}

   \keywords{Galaxy: kinematics and dynamics-- 
            Galaxy: evolution--
                Galaxy: disk --
                Galaxy: structure--
                Galaxies: interactions--}

   \maketitle
%


\section{Introduction}\label{sec:intro}

The snail shell or phase spiral that appeared in the vertical projection of the phase space \citep[][A18 hereafter]{Antoja2018} in the data of the second data release of the \Gaia mission (DR2; \citealt{Prusti2016,Brown2018}) informed us about a process of phase mixing that began following  a perturbation from which the Milky Way (MW) is still recovering. These findings provided further details about the vertical asymmetries discovered by  \citet{Widrow2012} and \citet{Williams2013}. 

Later we learned that the phase spiral shrinks in the $Z$ direction when moving from the outer to the inner parts of the Galaxy \citep{Laporte2019,Wang2019} as expected, that it is more prominent for cold orbits \citep{BlandHawthorn2019,Li2020}, that it shows mild changes with azimuth (e.g. slightly different density of stars along the phase spiral; \citealt{BlandHawthorn2019}), and that it is present in different age ranges \citep{Tian2018,Laporte2019,BlandHawthorn2019}, even in samples younger than 0.5 Gyr. The fanning of the phase spiral when colour coded according to the azimuthal velocity can be explained by the different vertical frequencies at different angular momenta (A18).

As for the causes of the phase spiral, most studies favour the hypothesis that it is due to the approach of the Sagittarius dwarf galaxy (A18, \citealt{Binney2018,Laporte2018,BlandHawthorn2021}, \citealt{Banik2022}). However, there are several complex aspects involved that remain to be fully understood, such as the effects of self-gravity discussed by \citet{Darling2019}, the mass of Sagittarius needed to activate phase spirals \citep[e.g. see discussions in][]{GarciaConde2022,Bennett2022}, or which Sagittarius pericentre/s excited the phase spiral.
Alternatively, the phase spiral found in the data of DR2 might have been caused by bending waves sparked by the buckling of the bar \citep{Khoperskov2019} or the impact of several (dark) subhalos \citep{Chequers2018,Tremaine2023}, and its formation and evolution may have been  helped by the wake of the halo excited by satellite passages \citep{Grand2022}. 

The shape of the phase spiral offers a way to constrain both the perturber and the potential of the Galaxy through the  vertical orbital frequencies of the stars. In A18, we assumed a set of frequencies of the MW and derived a perturbation time of 300-900 Myr. Ideally, both the potential and the perturbation times can be fitted at the same time as in \citet{Widmark2021}, although these two quantities appear to be degenerate in the fit.

In addition, the phase spiral alerted us to the fact that the assumption of a Galaxy in equilibrium may have biased certain determinations, such as that of the quantity of dark matter through the Jeans equations \citep{Haines2019, Chrobakova2020}. Furthermore, the vertical perturbations that led to the phase spiral might have other important consequences for the Galaxy. For example, they may have come accompanied by disturbances in the other components of phase space and/or changes in the morphology of the MW \citep{Laporte2018,BlandHawthorn2021,Antoja2022}, as proposed earlier, for example, by \citet{Younger2008} and \citet{Purcell2011}. The event might also be related to different episodes of strong star formation in the disc discovered with \Gaia data as well \citep{Mor2019,RuizLara2020}.

The publication of the third \Gaia  data release (DR3; \citealt{Vallenari2022}) offers a new opportunity to explore the phase spiral across the disc and its stellar composition through a better selection of populations. For example, \citet{RecioBlanco2022} showed that the phase spiral also appears when painted by metallicity, because of the correlation between metallicity and angular momentum. Recently, \citet{Hunt2022} discovered double spirals in the inner parts of the Galaxy and demonstrated that they could be due to breathing modes excited by internal structures such as the bar (but not necessarily a buckling bar) and/or by an external perturber, in a similar way to the double spirals from their simulations \citep{Hunt2021}.

Here, we explore the phase spiral with \Gaia DR3 and try to carry out a new determination of the perturbation time. In \s\ref{s_data}, we describe the data selection and corrections. In \s\ref{s_res}, we explore the local phase spiral, now with many more stars and with measurements of higher precision (\s\ref{s_loc}); we describe the method we use to detect the phase spiral and obtain its exact position in the $Z$--$V_Z$ plane (\s\ref{s_par}); and we further explore its 
spatial variations as functions of radius and angular momentum (\s\ref{s_LZ} and \ref{s_Zc}) as well as azimuth (\s\ref{s_phi}). In \s\ref{s_fit}, we provide a fit to the time of perturbation in a similar way to A18 but now considering the phase spiral at different angular momenta, which reveals interesting inconsistencies. In \s\ref{s_con}, we discuss our findings and present our conclusions.

\section{Data}\label{s_data}

We used data from \Gaia DR3 \citep{Vallenari2022}. We selected stars with available radial velocity and applied the following quality selections: (1) Astrometric quality selection: $\mathtt{RUWE}<1.4$, (2) selection in parallax quality:
$\mathtt{parallax\_over\_error}> 5$, and (3) selection of non-spurious solutions \citep{Rybizki2022}: $\mathtt{fidelity\_v2}> 0.5$.

We used distances from StarHorse \citep[hereafter SH,][]{Anders2022}  by default. We also used the photogeometric distances from \citet[][BJ hereafter]{BailerJones2021} for comparison (finding no important differences in our results) and also to test the robustness of the detections (see \s\ref{s_LZ}). In addition, we applied the following correction to the line-of-sight velocity for stars with $\mathtt{grvs\_mag} \ge 11$ and $\mathtt{rv\_template\_teff} <8500$ \citep{Katz2022}:
\begin{equation}\label{eq_vlos_correction}
\begin{split}
        \mathtt{radial\_velocity}  =   \mathtt{radial\_velocity} 
       \\ -0.02755*\mathtt{grvs\_mag}^2 \\+0.55863*\mathtt{grvs\_mag} -2.81129
\end{split}
.\end{equation}
As discussed in \citet{Blomme2022}, a correction is also necessary for the stars with $8500 \le \mathtt{rv\_template\_teff} \le 14\,500$~K \&  $6 \le \mathtt{grvs\_mag} \le 12$ but after the correction
 there is still a residual bias of a few $\kms$. As these stars are not very numerous, in our samples, we keep only stars with $\mathtt{rv\_template\_teff}<8500$ K. 
 
After all the above selections, the final sample with SH distances contains 25\,397\,569 stars, while that with BJ distances contains 26\,407\,121 stars. The queries used to retrieve the data from the \Gaia{} archive are presented in Appendix~\ref{app_queries}. In Appendix~\ref{app_dist}, we compare the two sets of distances and in Appendixes~\ref{appb_biases} and \ref{app_sf} we discuss the effects of distance errors and biases and the selection function, respectively, on the detected phase spiral.

We transformed the \Gaia observables into usual cylindrical phase-space coordinates. To this end, we used $R_0=8.277$ kpc \citep{Gravity2022}, $Z_\odot = 0.0208\,\kpc$ \citep{Bennett2019}, and $U_\odot=9.3$, $V_\odot+V_c(R_0)=251.5$ and $W_\odot=8.59\,\kms$ from the combination of the proper motion of SagA* from \citealt{Reid2020} and its radial velocity from \citet{Gravity2022}, where $U,V,$ and $ W$ are the usual heliocentric Cartesian velocities and $V_c(R_0)$ is the circular velocity curve at the position of the  Sun. Our reference system is a right-handed one with $\phi<0$ in the sense of rotation, and therefore also $V_\phi<0$ for most of the stars. We set the origin at the azimuth of the Sun ($\phi_\odot=0$). 

\begin{figure*}
   \centering
  \includegraphics[width=0.8\textwidth]{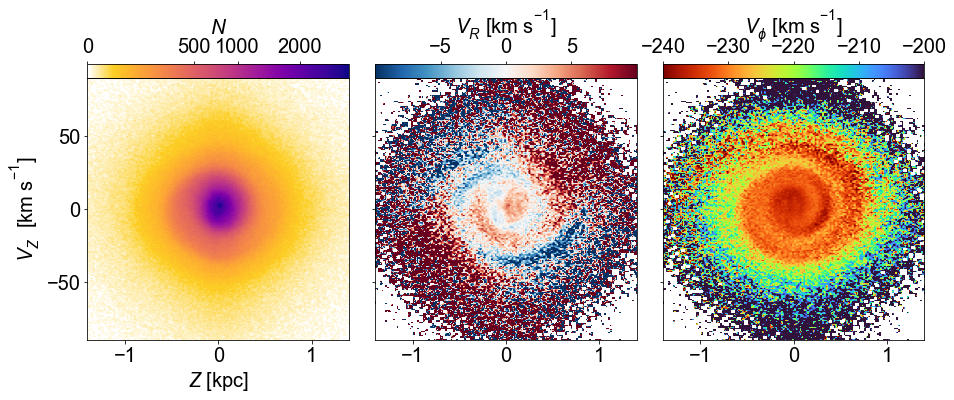}
   \caption{Phase spiral in the solar neighbourhood with \Gaia DR3 data. \emph{Left}: Two-dimensional histogram of the vertical projection of phase space ($Z$–$V_Z$) 
from the sample of stars with Galactocentric radii of
$R_\odot \pm 0.1$ kpc and $\phi=0\pm20\deg$. We used bins of
$\Delta Z = 0.02$ kpc and $\Delta V_Z=1\kms$. \emph{Middle}: Same projection but colour coded according to  median radial velocity $V_R$. \emph{Right}: Same projection but colour-coded according to  median azimuthal velocity $V_\phi$, adjusting the colour bar limits so as to maximise the appearance of the spiral.}
              \label{f_spiSN}%
\end{figure*}

\section{The phase spiral in DR3}\label{s_res}

\subsection{The local phase spiral}\label{s_loc}

In \f\ref{f_spiSN} 
 we look at the local phase spiral in the $Z$--$V_Z$ plane  by selecting stars with $R_0 -0.1\, \kpc <R<R_0 +0.1$ kpc and $-20 <\phi<20\deg$. This large range in $\phi$ is justified by the phase spiral not changing significantly with $\phi$ compared to $R$, as we see below. In A18, with DR2 data,  only the radial cut was performed, making a sample of 935\,590 stars. Here, we have a local sample with 2\,328\,004 stars. We note that we have increased the range in $Z$ and $V_Z$ of the panels in \f\ref{f_spiSN} compared to our previous work. We used bins with a size of $\Delta Z = 0.02$ kpc and $\Delta V_Z=1\kms$. The three panels of \f\ref{f_spiSN} (counts and coloured by median $V_R$ and by median $V_\phi$, respectively) now show a sharper signal. For example, the middle panel presents defined spiral segments of different $V_R$, including two blue segments in the upper left part. In the right panel, the phase spiral has a secondary branch at 50 $\kms$. While this could be partially observed in DR2 (A18, \citealt{Laporte2019}), now it is seen to be clearly separate from the other (main) spiral. This branching now reaches $Z=-1.2\, \kpc$, that is, between about three and five scale heights of the thin disc \citep[taken to be 220–450 pc,][and references therein]{BlandHawthorn2016}. 

\subsection{Detection of the phase spiral}\label{s_par}

\begin{figure}
   \centering
   \includegraphics[width=1.\columnwidth]{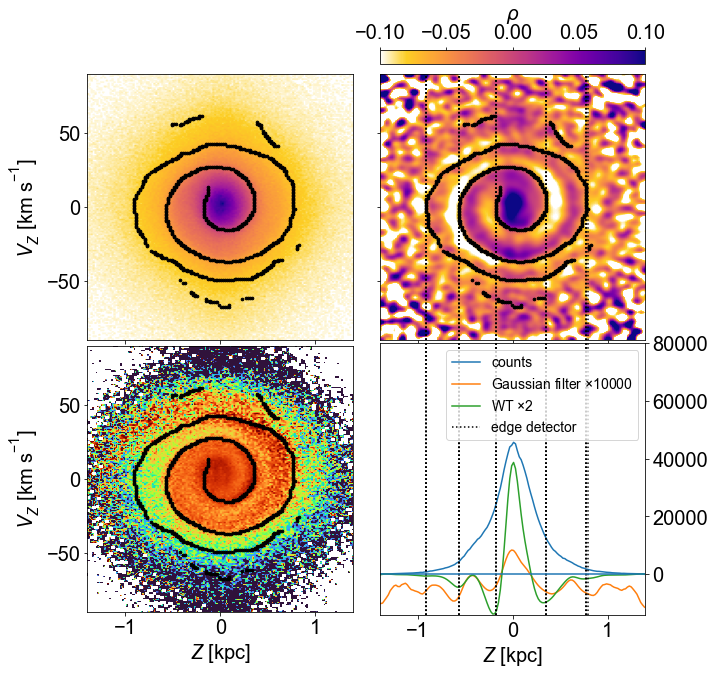}
   \caption{Phase spiral detected by the edge detection. \emph{Top left}: Output of the algorithm applied to the same selection of stars as in \f\ref{f_spiSN}, superposed to the counts from which it was computed. We used $\sigma=3$. \emph{Bottom left}: Same but superposed to the projection coloured according to $V_\phi$. \emph{Top right}: Comparison with the Gaussian filter (colour, see text). \emph{Bottom right}: Comparison with the Gaussian filter and the WT, with the respective values scaled to fit in the same plot, as indicated in the legend, in a one-dimensional case (see text).}
    \label{f_edge_detector_example}%
\end{figure}

In this section, we describe the method that we used to detect and study the shape of the spiral. It is based on an edge-detector algorithm applied to the counts in the $Z$--$V_Z$ space. We used the implementation in {\it python} $\mathtt{feature.canny}$ \citep{Canny1986} from $\mathtt{skimage}$ \citep{vanderWalt2014}\footnote{\url{https://scikit-image.org/docs/stable/api/skimage.feature.html#skimage.feature.canny}}. This technique starts by taking the 2D histogram. In most cases starting with a histogram on the logarithm of the number counts worked better (this is our standard option unless stated otherwise).  Next, the method reduces the noise using a Gaussian filter kernel with a certain $\sigma$ (in pixel units). We used a $\sigma$ that was found empirically to provide good detections. For instance, for the solar neighbourhood and binning defined in the previous subsection, we used $\sigma =3$. For volumes at smaller angular momenta $|L_Z|$  explored in the sections below, a larger $\sigma$ worked better. In particular, we used $\sigma$=4.5 for $|L_Z|<1600\,\kms\kpc$ and $\sigma$=3.5 for $|L_Z|\geq1600\,\kms\kpc$. The differences when using different $\sigma$ are very small for most of cases, and, by construction, they will be within the error bars of the detections (\s\ref{s_fit}). Next, the algorithm calculates the gradients for each pixel (bin) through the horizontal and vertical Sobel operators, takes the pixels with the maximum gradients, which are always perpendicular to the edges, and further selects the edge bins by hysteresis thresholding. This consists of taking all pixels with gradients above the high threshold (set to 20\%) and also, recursively, all the pixels above the low threshold (set to 10\%) that are connected to an already taken pixel. All the pixels with gradients below the low threshold are discarded. In practise, the algorithm returns a numerical matrix with values of 1 at the detected edges and 0 otherwise.  We note that this detector sometimes has problems in detecting the inner parts of the spiral. However, this is counterbalanced by the simplicity and speed of the algorithm and the good overall detection.

An illustration of the performance of the edge detector for the local phase spiral is shown in \f\ref{f_edge_detector_example}. In the top left panel, we superpose the detected edge to the vertical phase-space density. The edge detector perfectly delineates the main spiral and also finds signs of the upper branch in the local phase spiral. There is also a residual detection of an edge at $V_Z\sim-70\,\kms$ but with a less clear counterpart in the $V_\phi$-coloured projection. We note that the edge detector obtained from the counts projection approximately follows the spiral in the projection coloured by $V_\phi$ (bottom left panel). 

We compared the results from the edge detector with previously used methods, namely the Gaussian filter and the wavelet transform (WT; \citealt{Antoja2015} and references therein). Our Gaussian filter is similar to that used by \citet{Laporte2019}, but here we use $\rho=H_2/H_4-1$, where $H_2$ and $H_4$ are the original histogram of counts smoothed by a Gaussian filter (implementation $\mathtt{gaussian\_filter}$ from $\mathtt{scipy}$, \citealt{2020SciPy-NMeth}) with $\sigma_{\rm G}=2$ and $\sigma_{\rm G}=4$ (in units of the pixel size, $\Delta Z$ and $\Delta V_Z$), respectively. For the WT, we use the implementation $\mathtt{signal.cwt}$ from $\mathtt{scipy}$ with a width of the mother wavelet (Mexican hat) of 3 pixels. The top-right panel of \f\ref{f_edge_detector_example} compares the edge detector with the usual Gaussian filter, while the bottom-right panel compares the three methods  for a 1D projection with stars of $|V_Z|<3\,\kms$. The determinations from the three methods are very similar but the edge detector, by construction, is better at detecting the caustic edge that we are specifically aiming to locate.

\subsection{The phase spiral with radius and angular momentum}\label{s_LZ}

\begin{figure*}
   \centering
   \includegraphics[width=1.\textwidth]{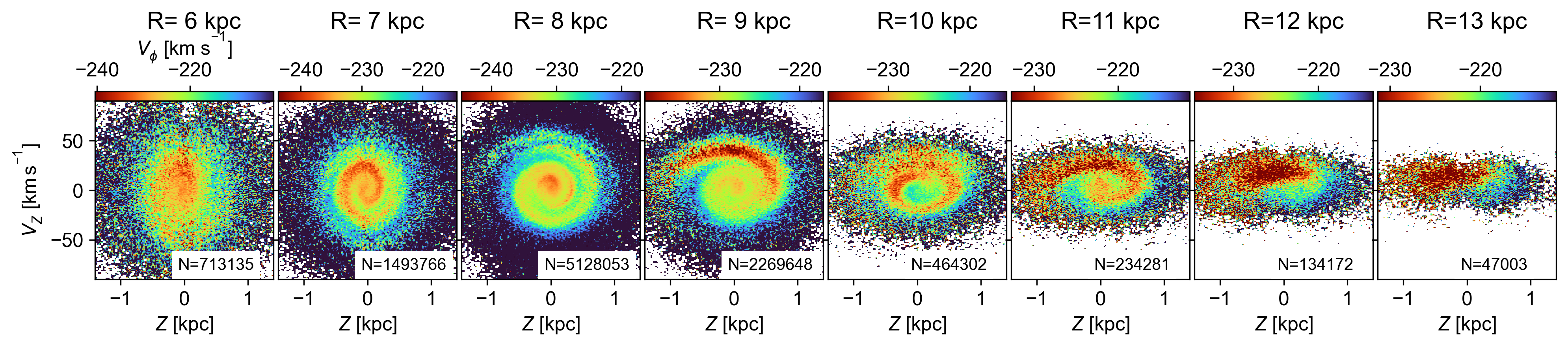}
  \caption{Phase spiral at different Galactocentric radii with \Gaia DR3 data. The panels show the vertical projection of phase space ($Z$–$V_Z$) coloured by median $V_\phi$. We only use stars with $|\phi|<20\deg$. The regions have a total radial width of 1 kpc and their centres are separated by the same amount. We used bins of
$\Delta Z = 0.02$ kpc and $\Delta V_Z=1\kms$. The colour-map ranges correspond to the percentiles 30 and 70 of the distribution of $V_\phi$ in each volume in order to maximise the appearance of the spiral. We indicate the number of stars in the bottom part of the panels.}
              \label{f_spiR}%
\end{figure*}
    
\begin{figure*}
   \centering
   \includegraphics[width=1.\textwidth]{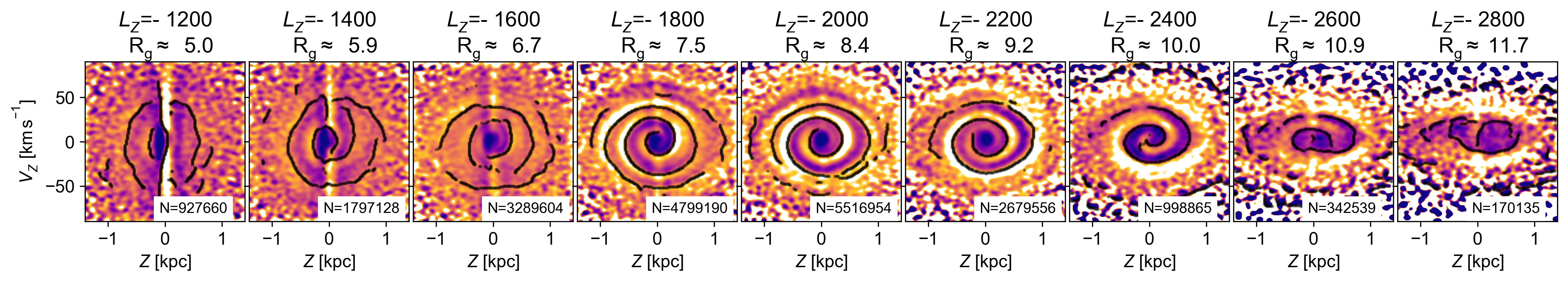}
\includegraphics[width=1.\textwidth]{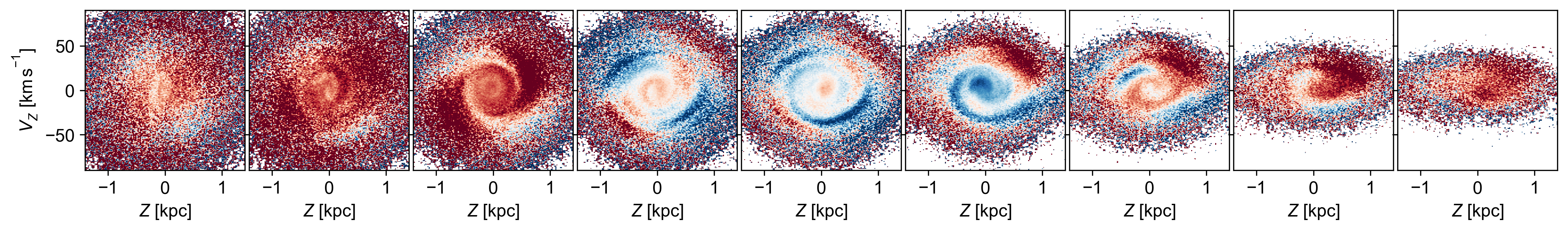}
   \caption{Phase spiral for different angular momenta with \Gaia DR3 data. We only use stars with $|\phi|<20\deg$. The selections have a total angular momentum width of 200 $\kms\kpc$ and their centres are separated by the same amount. We used bins of $\Delta Z = 0.02$ kpc and $\Delta V_Z=1\,\kms$. 
   \emph{Top:} Counts projection after Gaussian masking (see text for details). The colour bar is the same as in the top left panel of \f\ref{f_edge_detector_example}. We superpose our detected phase spiral via the edge detector method where we used $\sigma$=4.5 for $|L_Z|<1600\,\kms\kpc$ and $\sigma$=3.5 for $|L_Z|\geq1600\,\kms\kpc$. We indicate the number of stars in the bottom part of the panels. \emph{Bottom}: Projection coloured by $V_R$ with the same colour scale as in the central panel of \f\ref{f_spiSN}.}
              \label{f_spiLZ}%
\end{figure*}

In this section, we explore the spatial evolution of the phase spiral. First, in \f\ref{f_spiR}, we show the phase spiral coloured according to median $V_\phi$ for different Galactocentric radii, from inner (left) to outer radii (right), taking only stars with $|\phi|<20\deg$. The phase spiral becomes flatter with radius as already noticed before \citep[e.g.][]{Laporte2019,BlandHawthorn2019}. With DR3, we see that the phase spiral is detected from 6 to 11 kpc, thus extending beyond the previous radial limits. We also see hints of asymmetries reaching 5 and 13 kpc,  possibly limited by the selection effects there.  

We see a clear additional branching at $V_Z\sim50\,\kms$ in the volume at $R=8\,\kpc$, which is already seen in \f\ref{f_spiSN}, and now seems to extend slightly towards negative $V_Z$ as well. 
 From the animations\footnote{\label{note1}\url{https://teresaantoja.fqa.ub.edu/research/phasespiralDR3}} with more continuous sweeping in $R,$ we see that this arch or branch shifts to lower $V_Z$ with $R$, becoming mixed with the upper part of the main phase spiral and turning to an intense red ($V_\phi$ lower than the local median). We also detect what seems to be another turn of the phase spiral or faint arch at $V_Z\sim-50\,\kms$ at the region of 7 kpc. Curiously, we also see a pattern with a `hole' close to $(Z,\,V_Z)=(0,0)$ at 10 kpc (already noticed in \Gaia DR2, e.g. \citealt{Laporte2019}).

Separating by $L_Z$ instead of current radius yields a clearer spiral signal because binning in $L_Z$ groups stars with more similar vertical frequencies (see the Extended Data Figure 4 in A18 and the corresponding explanation), as was clearly demonstrated in \citet{Li2021}, \citet{Gandhi2022}, and \citet{Hunt2021}. However,  separating by angular momentum produces different biases, which are well explained by \citet{Hunt2022} for example, the most important being the dominance of eccentric orbits at the extreme angular momentum of the sample. 

In the top row of Fig.~\ref{f_spiLZ} we show phase spiral in counts for different bins of angular momentum. Again we only take stars with $|\phi|<20\deg$. 
In these panels an approximate measure of the guiding radius is given by simply doing $R_{\rm g}=L_Z/V_c(R_0)$, assuming a flat circular velocity curve. The black lines show the results of the edge detector (see caption for details). For the last three panels (large angular momentum $L_Z$) we applied the algorithm to the histogram of the counts (instead of the logarithm of the counts as in the other). This detects a better defined spiral for these cases but tends to detect a spiral biased towards inner parts. That is why we only use it for this figure but not when a quantitative comparison between panels needs to be done. In colours we plot the Gaussian filter as in the top right panel of \f\ref{f_edge_detector_example}.  
These panels show well defined phase spirals from $R_{\rm{g}}\sim$6 to 11 kpc, and some hints of them beyond these. In the left panels, we see a vertical band empty of stars that is explained by the lack of stars at low $|Z|$ towards the inner Galaxy due to extinction. The flattening of the spiral and their lower number of turns as one moves towards large $|L_Z|$ is evident. 
We note that the double spiral at small angular momentum discovered by \citet{Hunt2022} is not seen here. This could be due to a main phase spiral dominating and masking the double branches here. Alternatively, it could be that, since \citet{Hunt2022}  shows the phase spirals for a selection of nearby stars (1 kpc) split by angular momentum, the double spiral is only present for local stars with a small $|L_Z|$, that is when they are observed towards their apocentres. The upper separated branching seen at $R\sim8\,\kpc$ can be seen in the Gaussian filter from $R_{\rm{g}}\sim$7.5 to 10 kpc ($|L_Z|$ from 1800 to 2400 $\kms\kpc$). In some cases, for example at $|L_Z|=2200\,\kms\kpc$, further turns at larger $|Z|$ and $|V_Z|$ can be seen in the background colours.

In the bottom row of Fig.~\ref{f_spiLZ}, we colour code the vertical phase space according to median $V_R$.  A trend of the global $V_R$ with $L_Z$ is noticeable with redder colours (positive $V_R$) for small $|L_Z|$,  bluer (negative $V_R$) for intermediate $|L_Z|$, and again redder for large $|L_Z|$.  This is related to the $V_R$$-L_Z$ wave detected in \citet{Friske2019} (see also \citealt{Antoja2022}, and the middle panel of the fourth row in \f\ref{f_sf}). We also see a quadrupole at the outer parts of these diagrams, which is explained by the tilt of the velocity ellipsoid \citep{BlandHawthorn2019}. Aside from these, the phase-spiral segments appear clear and sharp. We note that the correspondence between density phase spiral and coloured by 
$V_R$ is not always direct. For example, the blue $V_R$ branch of the phase spiral at $L_Z=-2200\,\kms\kpc$ extends to $Z=1.4\,\kpc$ while the spiral segment in density continues to curl upwards. In some cases, we see  that $V_R$ changes its sign along the spiral segments ($L_Z=-2000\,\kms\kpc$).

\begin{figure}
   \centering
   \includegraphics[width=1.\columnwidth]{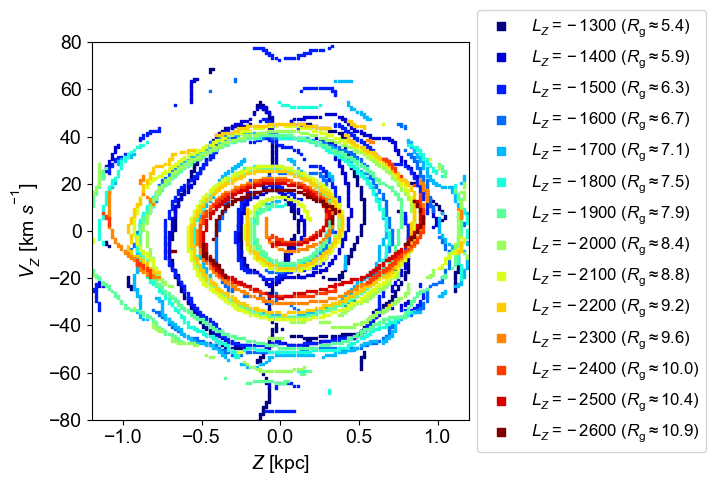}
   \caption{Superposed phase spiral for different angular momenta with \Gaia DR3 data. We select stars with $\phi\in[-20,20]\deg$. We use bins of $\Delta L_Z=200\,\kms\kpc$ and centres of bins separated by $100\,\kms\kpc$, from  $L_Z=-1300\,\kms\kpc$ (bluer lines) to $L_Z=-2500\,\kms\kpc$ (redder lines), as indicted in the legend. In parenthesis, we give an indicative measure of the guiding radius (in kpc), assuming a flat circular velocity curve. }
    \label{f_spiLZ2}%
\end{figure}

In Fig.~\ref{f_spiLZ2}, we superpose the phase spirals detected by our algorithm (\s\ref{s_par}) for different values of $L_Z$ as indicated in the legend, supplemented by their corresponding guiding radii $R_g$ assuming a flat circular velocity curve, selecting only stars at the azimuth of the Sun with $|\phi|<20\deg$. An animated version of the plot is available online\footnotemark[2]. 
We see an approximately continuous evolution of the phase spiral with angular momentum. The pitch angle increases as a function of $L_Z$ (or $R$, as already mentioned), as expected. The dark blue curves show some vertical lines related to the selection effects explained above. We note that for the same incremental value in $L_Z$, we do not see the same amount of morphological change; that is, for some angular momenta, the phase spiral changes more abruptly than for others (e.g. from dark to light blue).

\begin{figure}
   \centering
   \includegraphics[width=1.\columnwidth]{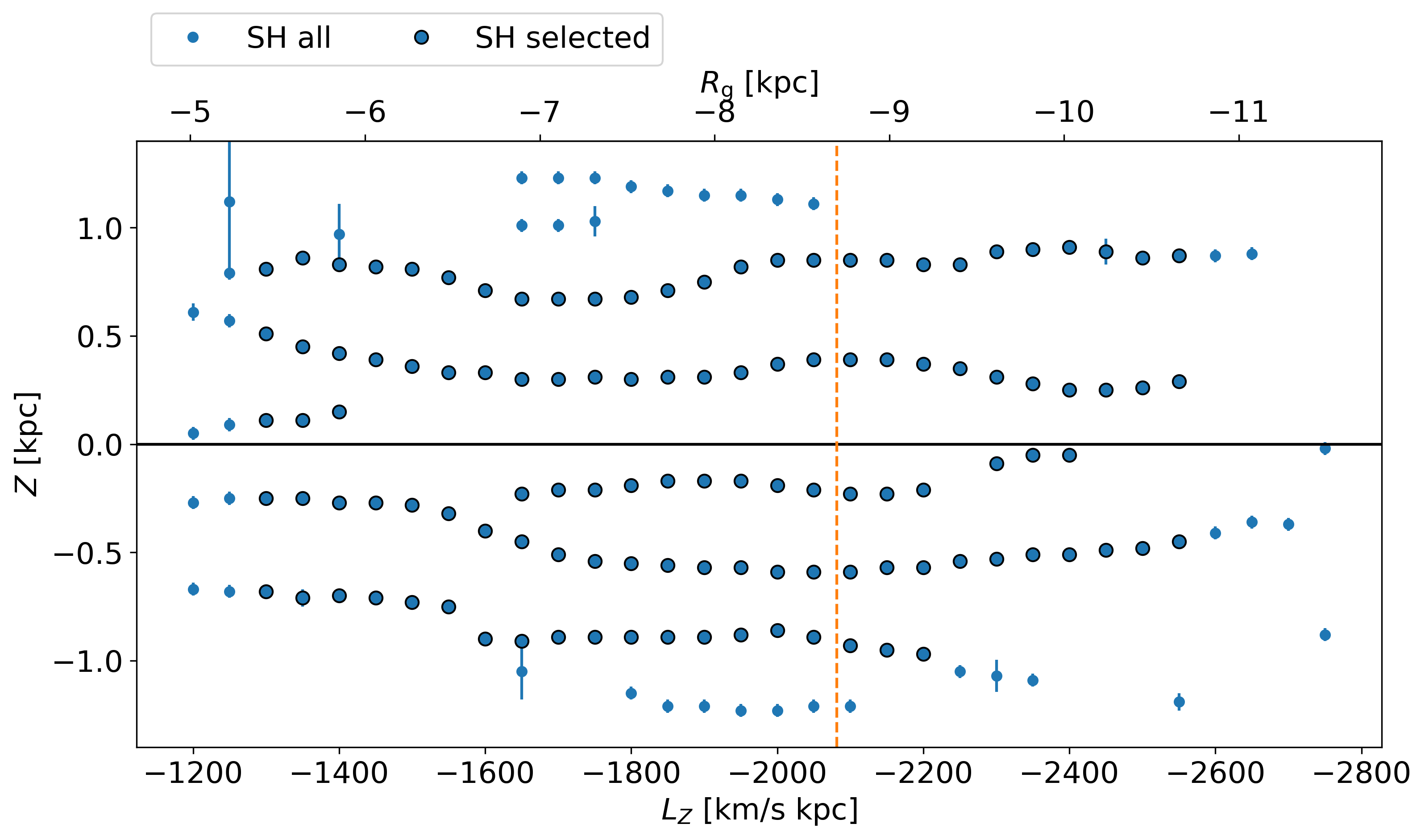}
   \caption{Crossing positions of the phase spiral (\zc). The blue circles and error bars mark the $Z$ coordinate for $V_Z=0$  of the phase spiral determined from the edge detector for the SH sample. We mark the positions used to infer the impact time with empty black circles. The dashed orange vertical line marks the angular momentum of the Sun.}
    \label{f_Zc}%
\end{figure}

\subsection{Crossing points of the phase spirals with $L_Z$}\label{s_Zc}

For the analysis of \s\ref{s_fit}, we need the $Z$ coordinate of the phase spiral when crossing the $V_Z=0$ line, which we call \zc{}. We determined these crossings using our edge-detection algorithm applied to the $Z$--$V_Z$ counts for different selections of angular momentum. We now use samples centred every 50 $\kms\kpc$ (therefore with more continuity) with a total width of $\Delta L_Z=100\,\kms\kpc$ (therefore with some overlap, and smaller than before in order to minimise variations of $L_Z$ within a bin). We considered only stars with $|\phi|<20\deg$. \ff\ref{f_Zc} shows \zc{} as a function of $L_Z$. 
In blue we show all the \zc{} from the edge detector, which show many continuous sequences corresponding to crossings that are well detected across several $L_Z$ but also noisy or spurious detections at the extreme $L_Z$, where data are less abundant and distance errors are larger.

Formally, we would assign an error to each \zc{} of the order of the bin size of the histogram in the $Z$ coordinate to which the edge detector is applied, that is 0.02 kpc. However, we see that using a slightly different $\sigma$ in the edge detector (from 2 to 5) may (in a few cases) produce differences that are around 0.03 kpc. We therefore take this as our baseline error. We also repeated the measurement of \zc{} with the BJ distances, computed the difference $\delta Z=|{\zcm}_{\rm{(SH)}}-{\zcm}_{\rm{(BJ)}}|$ between the crossings from SH and BJ (considering the closest points), and arbitrarily assigned an error of $\rm{MAX}(0.03\,\kpc,\,\delta Z) $ to each
\zc{}  from SH. Also, for the final analysis, we only considered \zc{} with errors smaller than 0.1 kpc (i.e. not considering points in SH that differ from BJ by more than 0.1 kpc). The comparison between the detections with SH and BJ is shown in \f\ref{f_dis}. The differences are in general very small, as expected from the small differences in distances at the ranges that we are probing (Appendix \ref{app_dist}). We also selected only points with $|Z|<1\,\kpc,$ since beyond this height the detections are also noisy and in some cases we are not certain whether they really correspond to a crossing of the phase spiral or to some edge of the global distribution. We only took points with $L_Z\in[-2600,-1300]\,\kms\kpc$ as the phase spiral is not clear beyond these limits  (\f\ref{f_spiLZ}). 

In \f\ref{f_Zc}, the points that we consider valid after all the  filters mentioned above are circled in black. 
At the intermediate $L_Z$ part of this figure, corresponding to angular momenta where there are more stars in our sample, more turns are detected (three for $Z<0$ compared to two at larger and smaller angular momentum), which is likely due to a combination of the selection effects and true differences in the phase spiral (expected larger pitch angle for larger $|L_Z|$, i.e. less turns). 
We also see some abrupt jumps such as at $L_Z\sim-1600\,\kms\kpc$ in the lowest turn ($\zcm\approx-0.8\,\kpc$). The influence of the selection function on the determinations of \zc{} is analysed in Appendix \ref{app_sf}, where we conclude that some of these jumps could be due to slight changes in the dominating population at each $L_Z$. 
 Aside from these jumps, the sequences of \zc{} show undulations. These are not expected from a simple interpretation of the phase spiral, in which we would see a continuous increase in |\zc| towards the outer parts of the Galaxy (large $|L_Z|$). This is because, in commonly used  Galaxy potential models, the gradient of vertical frequencies with vertical amplitude is smaller in the outer parts of the Galaxy (see Extended Data \f4 in A18), leading to a less tightly wound phase spiral. These undulations have important implications in the phase mixing times that we derive from these measurements in \s\ref{s_fit}.

\subsection{The phase spiral with azimuth}\label{s_phi}

\begin{figure}
   \centering
   \includegraphics[width=1.\columnwidth]{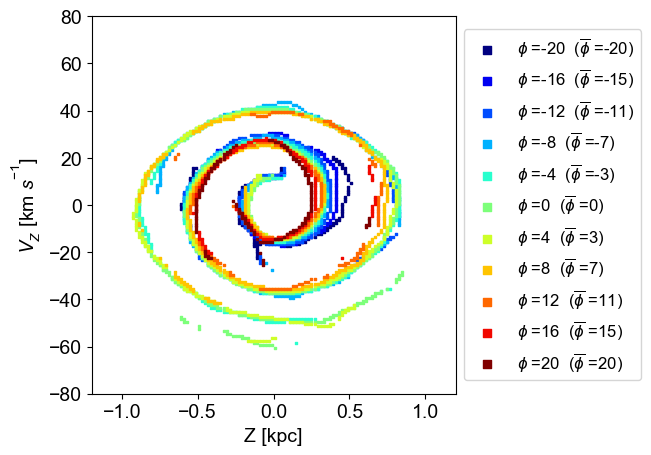}
   \caption{Superposed phase spiral for different azimuths with \Gaia DR3 data. We select only stars with $L_Z\in[-2100,-1800]$. We use bins of $\Delta \phi=4\deg$ and centres of bins separated by the same amount, from $\phi=-20  \deg$ (bluer lines) to $\phi = 20 \deg$ (redder lines), as indicted in the legend. In parenthesis, we give the average azimuth of the selection, which can differ by 1 $\deg$ compared to the centre of the selection in some cases. We used $\sigma=3.5$ for the edge detector.}
    \label{f_spiphi}%
\end{figure}

\f\ref{f_spiphi} shows the azimuthal variations of the phase spiral at fixed angular momentum. We choose  $L_Z\in[-2100,-1800]\,\kms\kpc,$ because this is the range around the value of $L_Z$ with maximum density in our sample. We see fewer changes with $\phi$ than with $L_Z$ but there is a clear gradual change.  In some parts of the phase spiral, there is greater variation with azimuth (see e.g. the turn at $Z\sim0.4$) than in other turns (e.g. the turn  at $-0.5\,\kpc$). In order to quantify these variations, we measure the slope of the crossings \zc{}  as a function of $\phi$ and find 
-0.006, -0.004, and 0.002 kpc/deg 
 for the crossings at 0.8, 0.3, and -0.6, respectively. The turns at $Z\sim-0.2$ and $Z\sim-0.9$ remain almost constant with azimuth. A simple phase difference with $\phi$ would produce the same level of fanning for each turn. Therefore, our observations require an extra deformation (e.g. different level of winding-up at different azimuth, i.e. different phase-spiral pitch angle) and/or strong population differences between positive and negative $Z$. Finally, we see that the lowest turn of the spiral at $V_Z\sim-50\,\kms$ is only seen in the azimuths close to the Sun, which is likely due to the increased number of stars in these volumes.

\section{Inferring the impact time}\label{s_fit}

\subsection{Method}\label{s_fit0}

\begin{figure*}
   \centering
   \includegraphics[width=0.7\textwidth]{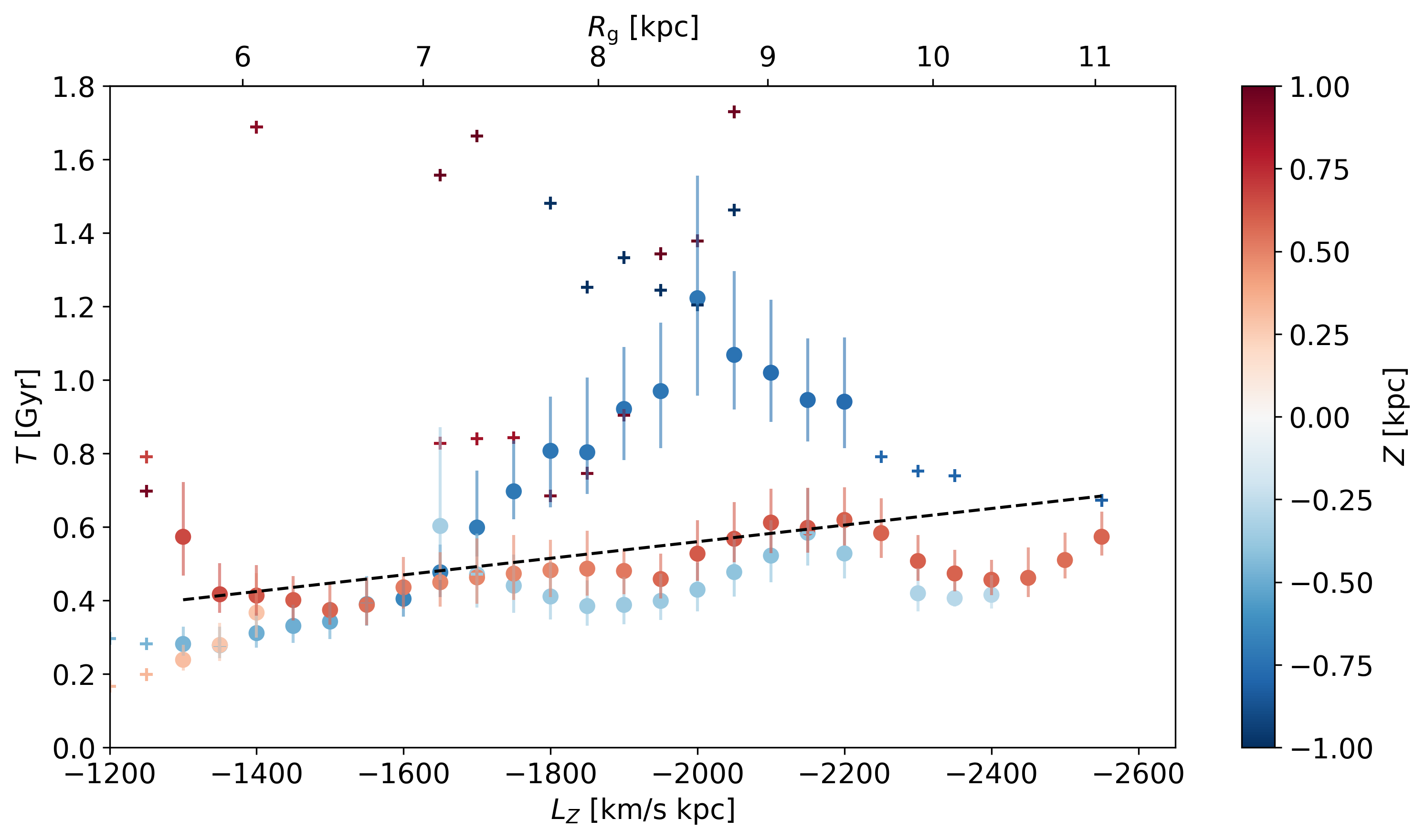}
   \caption{Time of the start of phase mixing as a function of angular momentum. The symbols give the time inferred from \e\ref{e_T} for the different consecutive crossing points of the phase spiral and are coloured by height above the plane. The circles with error bars are the points that we consider valid while plus signs are the remaining cases. An indicative guiding radius corresponding to the circular velocity curve of the McM17 potential is given in the upper horizontal axis. The dashed line marks a linear fit to the data points.}
    \label{f_T}%
\end{figure*}

\begin{figure*}
   \centering
   \includegraphics[width=0.7\textwidth]{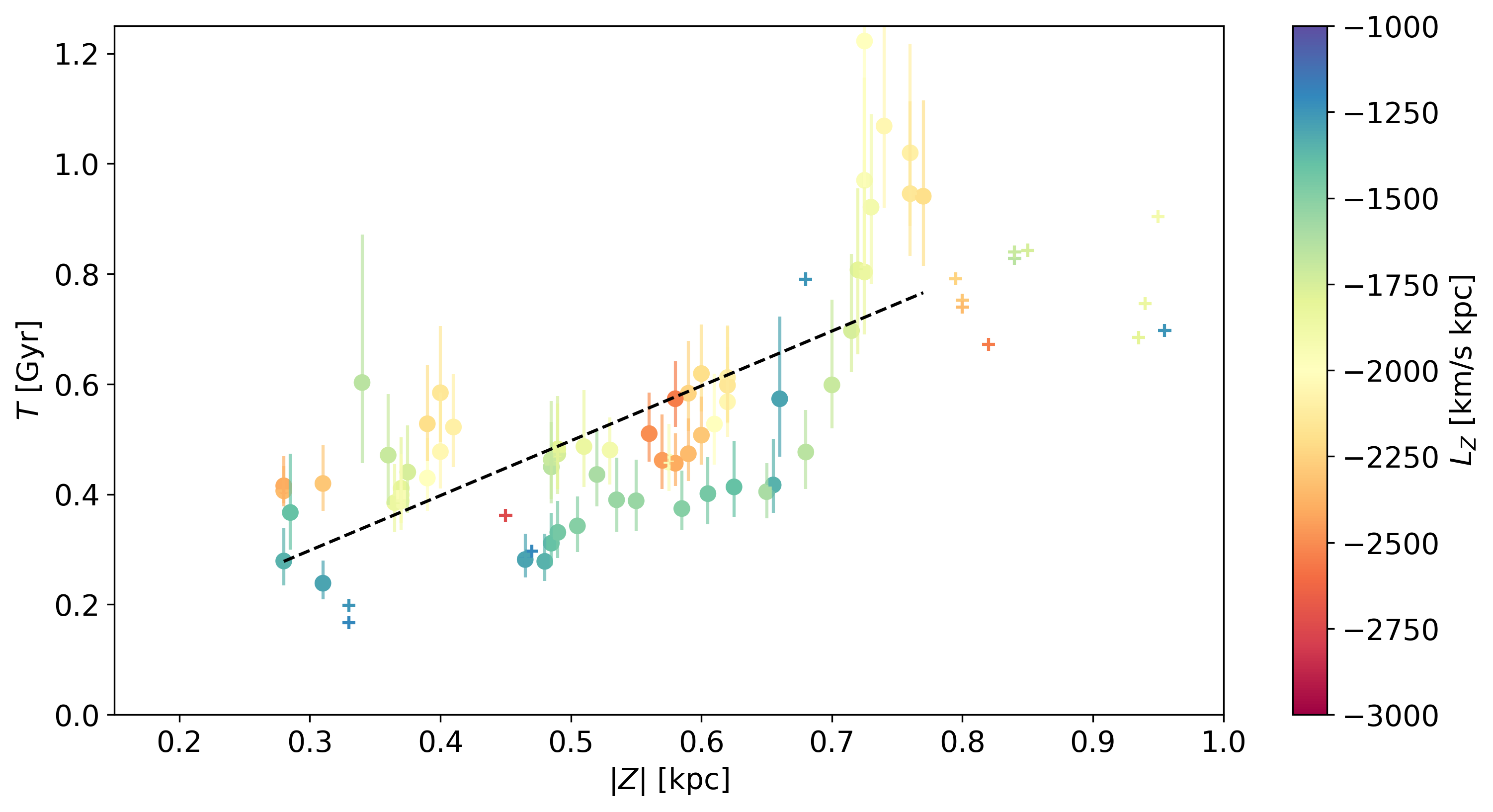}
   \caption{Time of the start of phase mixing as a function of height above the plane. The symbols give the time inferred from \e\ref{e_T} for the different consecutive crossing points of the phase spiral and are coloured by angular momentum. The circles with error bars are the points that we consider valid while plus signs are the remaining cases.  The dashed line marks a linear fit to the data points. }
    \label{f_TZ}%
\end{figure*}

In a phase-mixing process, the phase spiral gets more tightly wound with time and we can\ `rewind' it to infer the onset time of the phase-mixing process, which can then be linked to the time of the perturbation from which it originates. In A18, we used the consecutive turning points of the local phase spiral to constrain the impact time assuming a model of the MW potential from which the vertical frequencies were derived empirically. Here, we repeat the process of A18, but use data of the phase spiral at different angular momenta, helped by the increase in data at different positions across the Galaxy with the new \Gaia DR3. 

For this analysis, we use the crossings \zc{} of the phase spiral at $V_Z=0$ from the previous section (\f\ref{f_Zc}). Stars that have $V_Z=0$, such as in these crossing points, are currently at their maximum vertical height (above or below the plane):  $Z=\pm Z_{\rm{max}}$.  Therefore, we can estimate the vertical frequencies of these points by interpolating into a grid of frequencies computed as a function of $L_Z$ (or $R_g$) and $Z_{\rm{max}}$. Each pair of consecutive crossings\footnote{Here we refer to consecutive crossings for pairs of \zc{} that have either positive or negative $Z$. However, we note that one could also use pairs of \zc{} with different signs and change $2\pi$ for $\pi$ in \e\ref{e_T}. We examine this in \app\ref{app_test}.} ($Z_{\rm{c_1}}$, $Z_{\rm{c_2}}$) is then separated by a phase $2\pi$ and, thus, we can infer the impact time by their difference in vertical frequency:
\begin{equation}\label{e_T}
    T=\frac{2\pi}{\nu(L_Z,Z_{\rm{c_1}})-\nu(L_Z,Z_{\rm{c_2}})}.
\end{equation} 
Assuming that a single perturbation caused the phase spiral across the disc, this simple analysis should yield the same perturbation time at all $L_Z$ and for each pair of crossings. We note that with DR3 we are able to use more pairs of crossings (two in the $Z<0$ part of the spiral and one in the $Z>0$ part) at an angular momentum closer to that of the Sun, while in A18 we could only use two pairs in total. In A18, we used the crossings detected in the vertical phase space coloured by $V_\phi$ and here we use the crossings from the edge detector on the counts.

To build the grid of frequencies, we used {\tt AGAMA} \citep{Vasiliev2019} and simulated more than 8,000 orbits at $Z\geq0$ (and use the vertical symmetry of the potential for the $Z<0$ part) in the \citet{McMillan2017} potential (McM2017). This is our fiducial potential model. The orbits were integrated for about 50 revolutions. They all started with null $V_Z$ (i.e. at the top of their vertical oscillation). The other components of the initial conditions were set as follows: $\phi = 0^\circ$, $V_R = 0\,\kms$ (equivalently radial action $J_R=0$, but see below for orbits with non-null radial action). We then run through $Z$ (from -4$\,\kpc$ to 4$\,\kpc$ every 50~pc) and $L_z$ (from 500$\,\kms\kpc$ to 3000$\,\kms{}\kpc$ every 25$\,\kms{}\kpc$), 
numerically solving the implicit equation 

\begin{equation}\label{eq_Lz}
    L_z-\sqrt{-\frac{\partial\Phi(R,Z)}{\partial R}R^3} = 0
\end{equation}

\noindent in order to obtain the $R$, and thus $V_\phi=L_Z/R$, of the corresponding circular orbit. Subsequently, we sample each orbit with 20,000 points\footnote{We tested the sensitivity of the measured frequency to both the number of samples used and the number of revolutions, finding a negligible dependence on the former (as long as there are more than a few hundred points) and an asymptotic behaviour as a function of the latter. For 50 revolutions, the error on the measured frequency tends to be of the order of less than 1\%.} and measure the vertical frequency with a fast Fourier transformation \citep{FTT} of the signal in $Z$ as a function of time by taking the frequency of the dominant peak (multiple peaks can appear for very eccentric orbits with large vertical-amplitude oscillations). Below we explore other potential models and frequencies for different kinetic temperatures.  

\subsection{Time results}\label{s_fit1}

Interpolating in the above-mentioned grid of frequencies and using \e\ref{e_T}, we infer the time of the start of the phase obtained from each pair of crossings \zc{}. \f\ref{f_T} shows the results as a function of $L_Z$ (horizontal axis) and $Z$ (colour scale), where for the plot, we  used the average $(Z_{\rm{c_1}}+Z_{\rm{c_2}})/2$ of the points of the pair.  The time determinations from the \zc{} that we do not consider valid are marked with small plus symbols  (see text in \s\ref{s_LZ}; small blue circles in \f\ref{f_Zc}). These appear randomly scattered around the plot (many of these points are outside the vertical range). The circles with error bars are the time determinations from the valid \zc{} after the considerations of \s\ref{s_LZ}. The error bars\footnote{To compute the upper (lower) limit in each derived time, we used the limits of the errors of \zc{} that minimise (maximise) the frequency difference. That is, for pairs of crossings with $Z>0$, we  considered the lower limit of the smaller \zc{} and the upper limit of the larger \zc \ to obtain the lower limit on the time (maximum vertical-frequency difference). Conversely, the upper limit on time is obtained by combining the upper limit of the smaller \zc{} and the lower limit of the larger \zc.} are not symmetric because of the dependency of the frequencies on $Z$. 

 The valid times in \f\ref{f_T} (error bars) show different coherent sequences. The sequences from small $|Z_{\rm{c}}|$ (both positive and negative, lighter colours) are distributed around 0.5 Gyr and in many cases are consistent within the errors at a fixed $L_Z$. However, these points show increasing estimated times with $|L_Z|$. In addition, the darker blue points corresponding to the second crossing at $Z<0$ are at significantly larger times and also show a tendency to increase with angular momentum. Indeed, a linear fit with $L_Z$ to all the valid times 
 (dashed line in \f\ref{f_T}) illustrates this global trend.
 
 However, we see several oscillations in the different sequences, some of them corresponding to the oscillations in the \zc{} (\f\ref{f_Zc}) already mentioned. This is examined further in Appendix~\ref{app_sf}, where we conclude that while some oscillations could be related to changes in the dominating populations, the global trends with $L_Z$ and $Z$ might require a different explanation, which we discuss in \s\ref{s_con}.

We note that there is certain ambiguity in linking the different \zc{} into coherent sequences in \f\ref{f_Zc}. For example, it is not clear whether the sequence at $Z\approx-0.25\,\kpc$ starting at $L_Z=-1200\,\kmskpc$ continues towards higher or lower $Z$ when reaching $L_Z=-1600\,\kmskpc$. However, the time determinations do not depend on this link to  neighbouring points, because they are done independently at each $L_Z$ from consecutive crossings in $Z$. There might be a problem if some crossings are missing but it is unlikely that we have missing crossings in between detected ones: the undetected crossings would be likely located at lower |Z| (the phase spiral is highly undetermined in the central parts of the $Z$--$V_Z$ diagram) and/or at higher |Z| (due to fewer counts). These missing crossings would not bias the estimated times from the intermediate crossings (i.e. would not change from the ones appearing in \f\ref{f_T}).

\f\ref{f_TZ} is equivalent to \f\ref{f_T} but shows the time of the start of phase mixing as a function of $|Z|$ (horizontal axis) and $L_Z$ (colour scale). A general trend of time increasing with $|Z|$ is observed, as seen in the linear fit 
(dashed line in \f\ref{f_TZ}). However, this fit would depend on the angular momentum (colours in \f\ref{f_TZ}). Part of these time discrepancies for different $Z$ at fixed $L_Z$ could be explained by the fact that the vertical frequency is not perfectly defined for eccentric orbits with large oscillations about the midplane. As we estimate the frequency numerically and take that of the dominant peak, we expect that part of this problem is  mitigated.

Ignoring the trend with $Z$ and $L_Z$ (which we discuss in \s\ref{s_con}) for now,  we obtain an average start of phase mixing time of 0.5 Gyr (0.4 Gyr for the error-weighted average, $w=1/\sigma^2$). We find that determinations are within $[0.3, 0.9]$ Gyr 80\% of the time. The variations with $L_Z$ and $Z$ are larger than the statistical error (error bars).  \t\ref{t_T} (first row) also provides other statistics such as the minimum and maximum values. 
%

\begin{figure}
   \centering
   \includegraphics[width=1.\columnwidth]{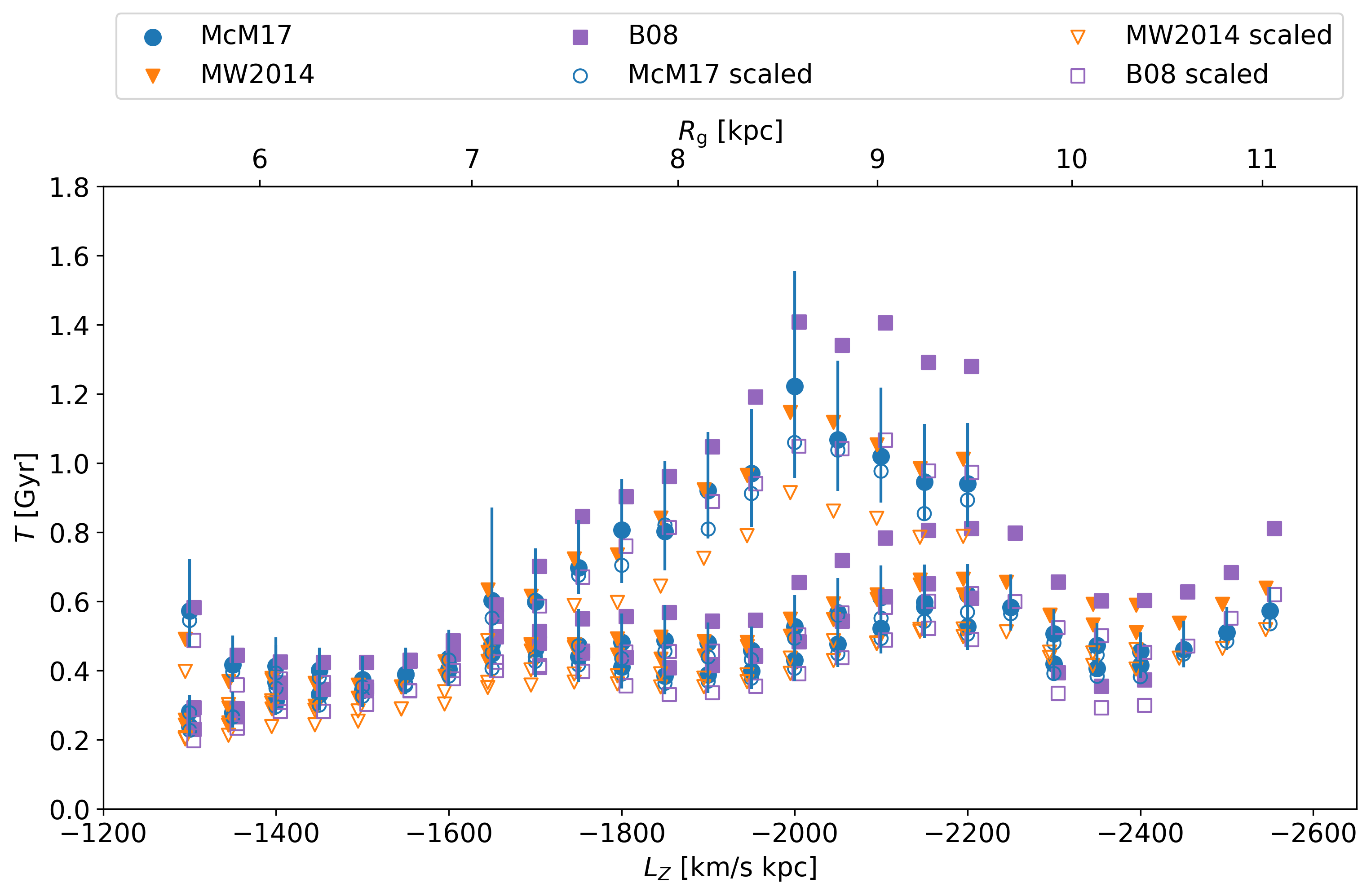}
   \caption{Time of the start of phase mixing for different potentials. The figure is similar to \f\ref{f_T} but we use frequencies of different potential models of the MW, as indicated in the legend.}
    \label{f_Tm}%
\end{figure}

\begin{figure}
   \centering
   \includegraphics[width=1.\columnwidth]{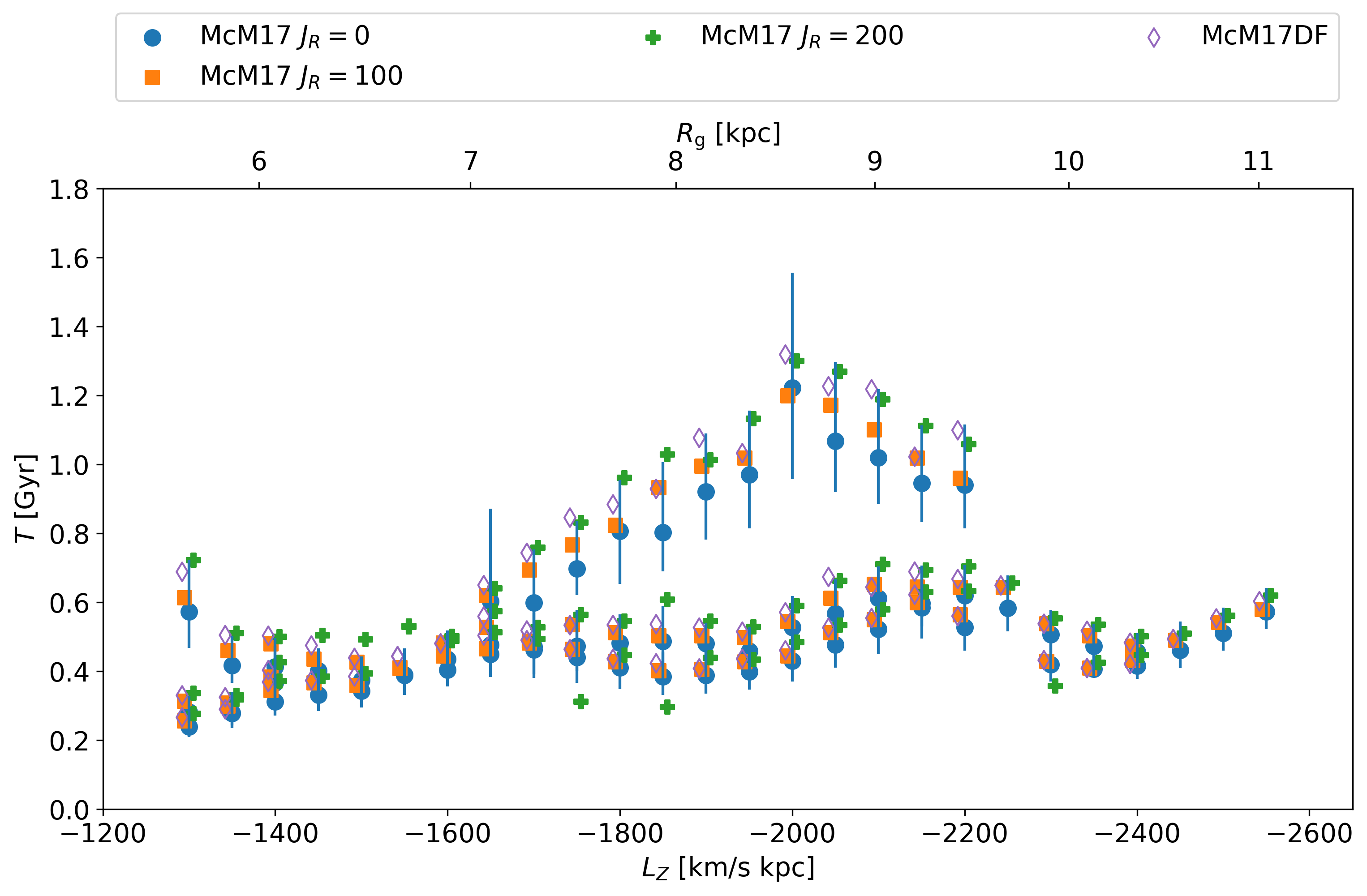}
   \caption{Time of the start of phase mixing for different kinetic temperatures. The figure is similar to \f\ref{f_T} but we use orbits with different radial actions and methods, as indicated in the legend.}
    \label{f_To}%
\end{figure}

\begin{figure}
   \centering
   \includegraphics[width=1.\columnwidth]{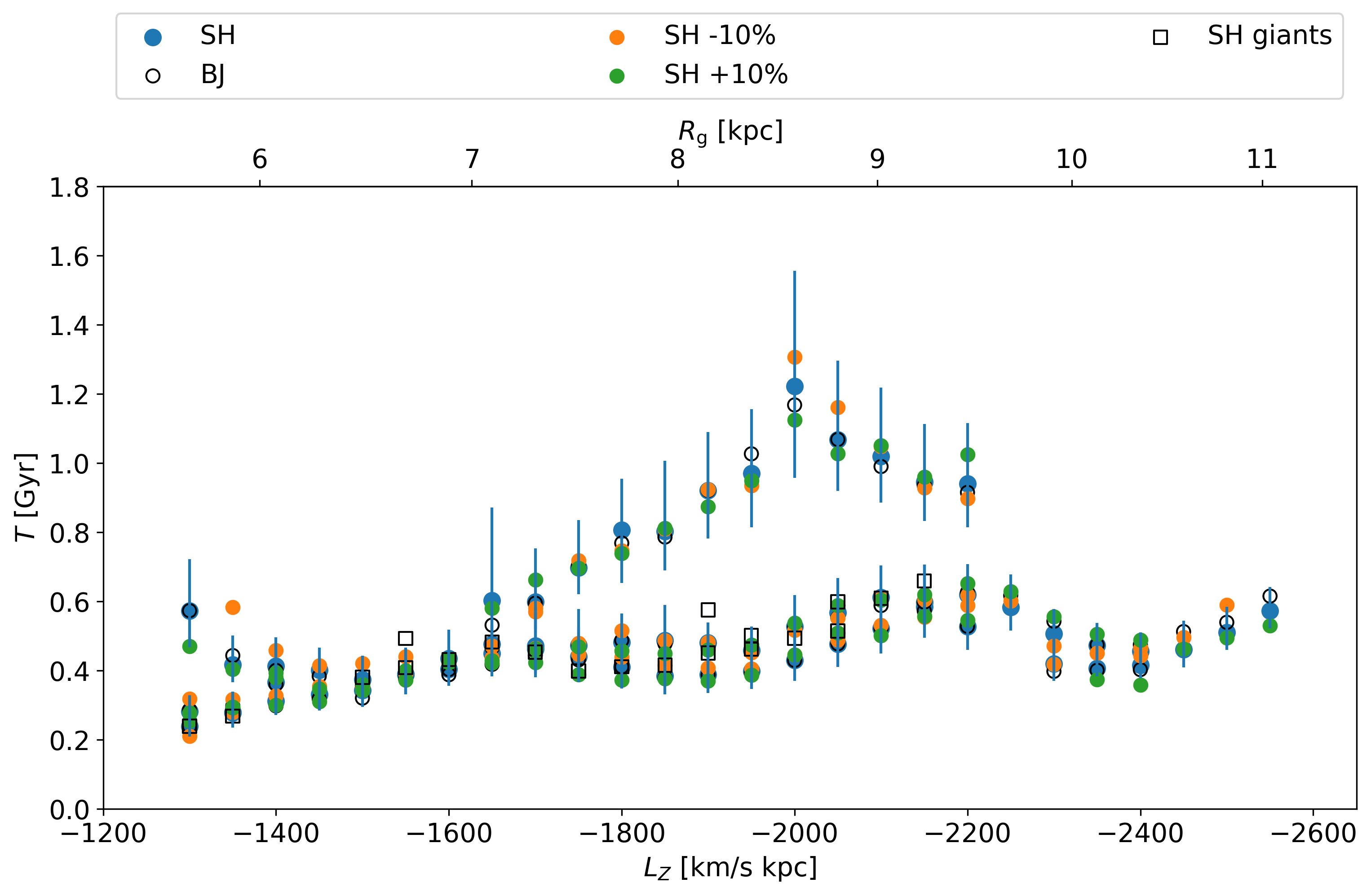}
   \caption{Time of the start of phase mixing for different distances and data. The figure is similar to \f\ref{f_T} but we compare our main results with results using distances from BJ, using distances from SH but adding a distance bias of $\pm10\%$, and using a sample of giant stars, as indicated in the legend.}
    \label{f_Td}%
\end{figure}

\begin{table*}
\caption{Phase-mixing times from the phase spirals. We show weighted and unweighted averages (columns 2 and 3) with their standard errors, minimum and maximum values (columns 4 and 5), and ranges enclosing 80\% of times (column 6). The last two columns (7 and 8) compare the times with the fiducial case (first row), showing the average differences and the ranges of differences. The different rows are for different cases, as explained in the text. For the times derived from the \vzc{} (last row), we do not compute the differences with the fiducial case as there is no point-by-point correspondence.}            
\label{t_T}      
\centering                          
\begin{tabular}{l c c c c c c c}        
\hline\hline                 
Model & mean($T$) & weighted mean($T$)& min($T$)&max  $T$&[P10, P90]$_T$&  mean($T-T_0$)& [P10, P90]$_{T-T_0}$\\    
\hline                        
McM17 $J_R=0$ &0.5 $\pm$ 0.03 &0.4 $\pm$ 0.01 &0.2 &1.2 &[0.3, 0.9] &- &-\\ 
MW2014 &0.6 $\pm$ 0.03 &0.4 $\pm$ 0.01 &0.2 &1.1 &[0.3, 0.9] &0.02 &[-0.04, 0.08] \\ 
B08 &0.6 $\pm$ 0.04 &0.5 $\pm$ 0.01 &0.2 &1.4 &[0.4, 1.0] &0.09 &[-0.00, 0.21] \\ 
McM17 scaled &0.5 $\pm$ 0.02 &0.4 $\pm$ 0.01 &0.2 &1.1 &[0.3, 0.8] &-0.03 &[-0.05, -0.01] \\ 
MW2014 scaled &0.4 $\pm$ 0.02 &0.3 $\pm$ 0.01 &0.2 &0.9 &[0.3, 0.7] &-0.09 &[-0.17, -0.03] \\ 
B08 scaled &0.5 $\pm$ 0.03 &0.4 $\pm$ 0.01 &0.2 &1.1 &[0.3, 0.9] &-0.03 &[-0.06, 0.02] \\ 
McM17 $J_R=100$ &0.6 $\pm$ 0.03 &0.5 $\pm$ 0.01 &0.3 &1.2 &[0.4, 1.0] &0.03 &[0.01, 0.07] \\ 
McM17 $J_R=200$ &0.6 $\pm$ 0.03 &0.5 $\pm$ 0.01 &0.3 &1.3 &[0.4, 1.0] &0.08 &[0.04, 0.15] \\ 
McM17 DF &0.6 $\pm$ 0.03 &0.5 $\pm$ 0.01 &0.3 &1.3 &[0.4, 1.0] &0.06 &[0.02, 0.12] \\ 
$d_{\rm{SH}-10\%}$ &0.5 $\pm$ 0.03 &0.4 $\pm$ 0.01 &0.2 &1.3 &[0.4, 0.9] &0.02 &[-0.03, 0.05] \\ 
$d_{\rm{SH}+10\%}$ &0.5 $\pm$ 0.03 &0.4 $\pm$ 0.01 &0.3 &1.1 &[0.3, 0.9] &-0.01 &[-0.05, 0.03] \\ 
$d_{\rm{BJ}}$ &0.5 $\pm$ 0.03 &0.4 $\pm$ 0.01 &0.2 &1.2 &[0.3, 0.9] &-0.00 &[-0.03, 0.03] \\ 
SH giants &0.5 $\pm$ 0.02 &0.4 $\pm$ 0.02 &0.2 &0.7 &[0.4, 0.6] &0.04 &[-0.02, 0.10] \\ 
SH $V_Z$ & 0.6 $\pm$ 0.04 &0.4 $\pm$ 0.01 &0.3 &1.7 &[0.3, 0.9] &- &-\\ 
\hline                               
\end{tabular}
\end{table*}

\subsection{Results for different models and data}\label{s_fit2}

\begin{figure}
   \centering
   \includegraphics[width=1.\columnwidth]{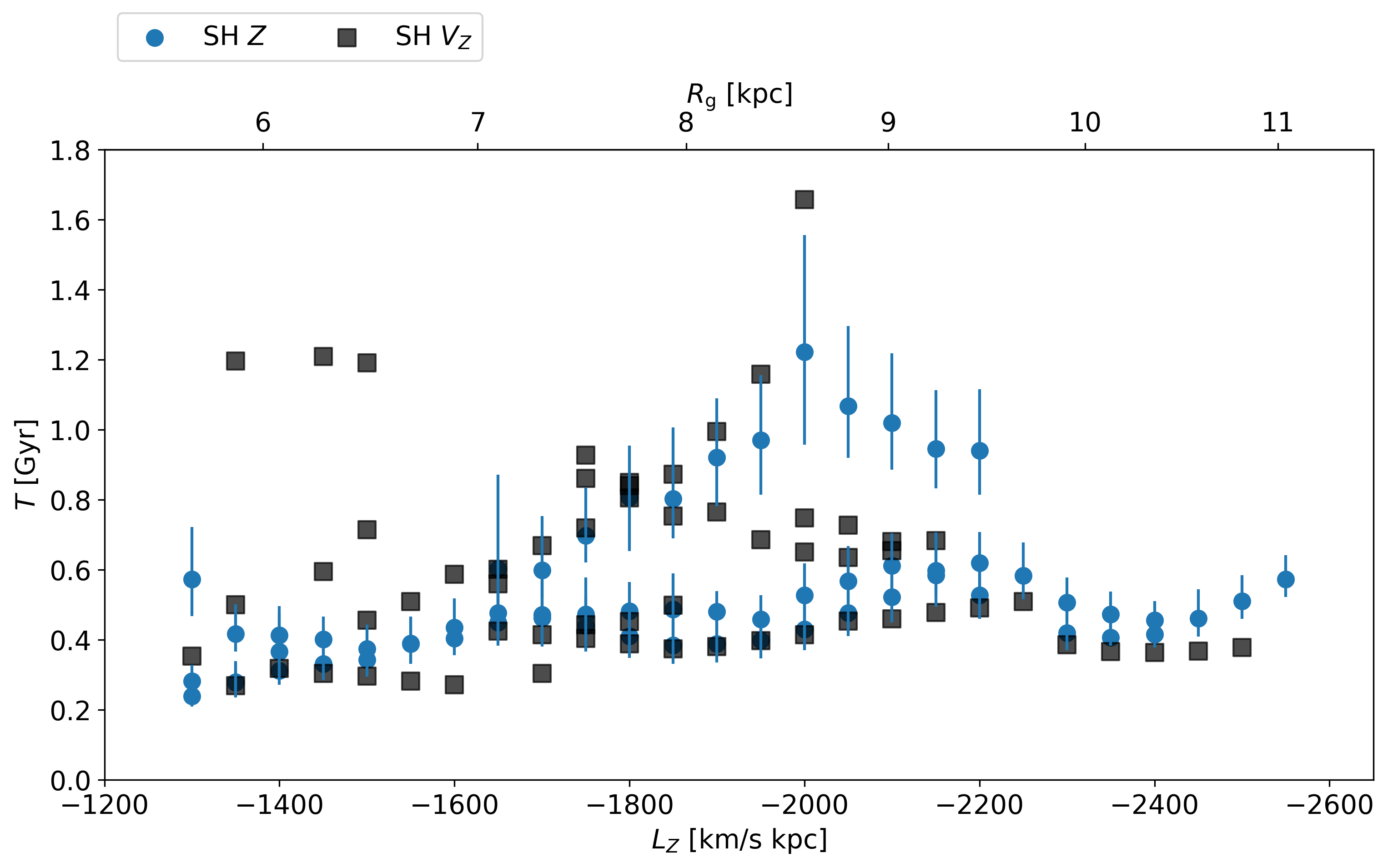}
   \caption{Time of the start of phase mixing for different parts of the phase spiral. The figure is similar to \f\ref{f_T} but we use the fiducial case (blue dots) computed from \zc{} but also from \vzc{} (grey squares) to derive the times, as indicated in the legend.}
    \label{f_TVZ}%
\end{figure}

We examined the dependency of the results on different potential models (\f\ref{f_Tm}), the frequencies for different kinetic temperatures (\f\ref{f_To}), and the possible systematic errors on distance (\f\ref{f_Td}), and looked at the effects of taking the $V_Z$ crossings of the spiral with the $Z=0$ axis instead of the \zc{} (see text below, \f\ref{f_TVZ}). In all these figures, blue circles show our fiducial case presented above, which consists of using the McM17 potential (non-scaled), SH distances, and $J_R=0$. The time statistics are given in Table \ref{t_T}, where the last two columns indicate the differences between the times of these new cases and the times from the fiducial case ($T_0$). In most cases, the times are within the error bars of the fiducial case. We find the largest differences when we change the assumed potential model for the MW. In particular, we obtain systematically larger times for Model I (here B08) from \citet{Binney2008} (on average 0.09 Gyr larger), and systematically smaller times for the MWpotential2014 (here MW2014) from \citet{Bovy2015} when scaled to fit the values of $V_c$ and $R_0$ used in the data (on average 0.09 Gyr smaller). We also obtain systematically larger times when using not-circular orbits: for example 0.08 Gyr larger times for orbits with radial action $J_R=200\,\kms\kpc$.

However, the main trends in the time determinations (in angular momentum and in height) remain the same in all cases. Our conclusions are therefore applicable to all the sets examined. Below we provide details of this analysis; the  conclusions are provided in Sect.5. 

We first explored different potential models (\f\ref{f_Tm}), whose circular velocity curves are compared in \f\ref{f_rc}. We show the gradient of the vertical frequencies of these models in \f\ref{f_fg}. The gradients are slightly different as a function of radius and height but yield only small differences in the time determinations. The orange triangles in \f\ref{f_Tm} use MW2014. This model returns slightly smaller times at smaller $|L_Z|$ and larger times at larger $|L_Z|$ but with null average differences when taking all individual time measurements. Purple squares are for B08. With this case, we obtain the largest differences from all the variations explored in this section, that is, systematically larger than our fiducial case and with several points at $2\sigma$. The average differences are of 0.09 Gyr and 80\% of the time measurements are within $[ 0.4,1.]$ Gyr. We also compared results with these potential models but now scaled to fit the values of $V_c$ and $R_0$ used in the data instead of the original parameters from the respective articles. To do this, we use the built-in scaling mechanism of {\tt AGAMA} by providing a mass scaling factor $(V_c(R_0)_{\rm desired}/V_c(R_0)_{\rm default})^2$ at the time of creating the gravitational potential, which results in the velocities and frequencies being scaled as the square root of this scaling factor. The results for the scaled potentials are shown with the empty symbols in \f\ref{f_Tm}. It is now the MW2014 (scaled) model that gives maximum differences with the fiducial case, yielding smaller times ($[0.3,0.7]$ Gyr).

   Secondly, we examined different orbital conditions (\f\ref{f_To}), using frequencies for different kinetic temperatures. In our fiducial case, we use orbits with $J_R=0$, that is, nearly circular orbits. Here, we increase the eccentricity of the orbits in two different ways and use the corresponding new frequencies. Frequencies for orbits with larger radial actions ($J_R=100$ \& $J_R=200\,\kms\kpc$, orange squares and green crosses in \f\ref{f_To}) yield larger times in general because hotter orbits have smaller differences between frequencies at different heights. Nevertheless, the average time does not change significantly. We also explored frequencies for the most populated part of the distribution function (DF; purple diamonds in \f\ref{f_To}). This is done by fixing $R$ and $Z$ and estimating the vertical frequency of the orbit with the highest probability in the DF at that location of space, that is, at $Z_{\rm{max}}$ ($V_Z=0$). With these different frequencies, we also obtain larger times, similar to the $J_R=200$ case.
       
    Thirdly, we compare our results with those of samples with different distances and different selections of stars (\f\ref{f_Td}).
    This is also examined in Appendix~\ref{appb_biases}. 
    Using the BJ distances, our results change only very slightly (empty black circles  in \f\ref{f_Td}). This is not surprising given the agreement between both sets of distances for our selected samples. When we consider possible distance biases, we see that if we decrease or increase the distances by 10\% (orange and green circles in \f\ref{f_Zc2}), mimicking a correction of overestimated or underestimated distances, respectively, there is an increase or decrease, respectively, in the value of $|Z_{\rm{c}}|$ by approximately the same amount  (Appendix~\ref{appb_biases}, \e\ref{eq_z_bias}), except at the extreme $L_Z$ where selection effects might be playing a role. 
    However, as \e\ref{e_T} uses differential frequencies, we see that the times obtained do not change significantly (orange and green triangles in \f\ref{f_Td}, consistent with the non-biased SH distances within the errors in most cases).
 We also repeated our analysis using the selection of giant stars from \citet{Drimmel2022} obtained from the parameters of the \Gaia{} \texttt{gspphot} pipeline with the same quality filters as for our main sample. The new \zc{} and derived times are shown in \f\ref{f_Zc2} and  \f\ref{f_Td} (empty black squares in both figures), respectively. In almost all cases, the new values fall within the error bars of the fiducial case and the trend in angular momentum remains the same. We cannot check whether the times also increase with $|Z|$ because only the inner parts of the spiral are detected due to the small number of stars in this selection (4\,901\,270).

 Finally, we also derive the times from the $V_Z$ crossings of the phase spiral at $Z=0$, which we name \vzc{}, in addition to the $Z$ crossings at $V_Z=0$ (\zc), as we did in  A18. More details are given in \app\ref{app_test}. The location of \vzc{}  with respect to the \zc{} crossings suggest, among other things, that the assumed potential to infer the frequencies (and therefore the impact times) is not fully correct. We return to this point in our discussion (\s\ref{s_con}). As for the obtained times, the overall numbers do not change when considering these new crossings (\f\ref{f_TVZ}).

\section{Summary, discussion, and conclusions}\label{s_con}

We examined the phase spirals in the MW with the new data from \Gaia DR3. Our findings can be summarised as follows.
\begin{itemize}
    \item We find a clear increase in the $Z$ and $V_Z$ ranges in which the phase spiral is detected. Surprisingly, we see the  phase spiral extending down to $-1.2\,\kpc$, which is well into the realm of the thick disc. This does not necessarily mean that it is made of thick-disc stars but could be made of thin-disc stars that have been largely displaced. 
    \item  We detect phase spiral turns that extend in $V_Z$ beyond $\pm 50\,\kms$, meaning that the perturbation produced a velocity kick at least of this amount and/or affected stars with these high vertical velocities. 
     \item An increase in the range of detection of the phase spiral in Galactic radius and azimuth is seen in the new DR3 data. We see clear phase spirals between 6 and 11 kpc in radius and also evident asymmetry, probably indicative of a poorly resolved spiral, down to 5 kpc and up to 13 kpc (\f\ref{f_spiR}). 
    \item We also see clearer phase spirals in counts and their coloured versions in radial and azimuthal velocities.
    \item  We detect a secondary branch of the phase spiral at large positive $V_Z$ in the counts projection and in the phase spiral painted by $V_\phi$. This is observed in the local phase spiral (see also A18 and  \citealt{Laporte2019}) but also at angular momenta between 1800 and 2400 $\kms\kpc$. This branch seems to merge with the well-defined main phase spiral moving towards the outer Galaxy. A somewhat similar branching is seen in the simulations by \citet{Hunt2021} (their figure 5). This secondary branch might be caused by the complexity of the phase-mixing process (e.g. different groups of actions, stemming from non-uniform initial conditions, or with strong dependence on the position of the disc) or different perturbations.
    \item The phase spiral has different morphology when splitting by angular momentum, including the expected flattening with $Z$ at larger $|L_Z|$ (also seen in DR2, e.g. \citealt{Laporte2019}, \citealt{BlandHawthorn2019}) but also trends departing from this expected flattening in a usual MW gravitational potential. While part of these trends could be due to selection effects (Appendix~\ref{app_sf}), a more straightforward explanation might be our overly simplistic modelling of the phenomena involved (see below). 
    \item We see differences in the spiral in counts and coloured by $V_R$ at certain angular momentum. As seen in the simulations in \citet{BlandHawthorn2021} for an impact with a galactic satellite, the vertical projection of phase space coloured by $V_R$ could be a result of the combination of: (1) the density wave induced after a perturbation (mostly a two-armed spiral  density structure in the impulsive and distant impact conditions; \citealt{Toomre1972,Struck2011}) that is linked to a $L_Z$--$V_R$ wave \citep{Antoja2022}, and (2) the vertical waves induced after the same perturbation. However, the exact way in which the planar and vertical disturbances are coupled is not yet well understood. Joining simple modelling of satellite perturbations such as those in \citet{Gandhi2022} (vertical) and \citet{Antoja2022} (planar) could help in this respect. Exploring more complex models with sufficient resolution, such as those in \citet{Hunt2022}, is also necessary.
    \item We detect mild changes of the phase spiral at different azimuth. Their Z coordinates at $V_Z=0$ can differ by up to -0.006 kpc/deg but these changes are not uniform (i.e. they depend on the turn of the phase spiral). 
\end{itemize}

Finally, we estimated the phase-mixing times from the phase spiral for the first time using data at different angular momenta. Our findings from this part of the analysis can be summarised as follows.
\begin{itemize}
    \item There is a large amount of variation in the times derived from measurements of the phase spiral at different heights (different turns) and at different angular momenta: we see an increase in time with $|L_Z|$ and especially with $|Z|$. The time differences can be of 0.7 Gyr using a single phase spiral at the angular momentum close to that of the Sun but different turns of the phase spiral (i.e. different $Z$) and of 0.4 Gyr considering different angular momenta.
    \item The average time is of 0.5 Gyr. For 80\% of our determinations, we find times in the range of 0.3-0.9 Gyr.
    \item We find slightly larger times when using the B08 model (0.4-1 Gyr), systematically larger times for hotter orbits (0.4-1. Gyr), and smaller times for a rescaled version of MW2014 (0.3-0.7 Gyr).
    \item The values obtained are very similar to the ones given in A18, which were 0.3-0.9 Gyr and are consistent in their upper limits with the determinations from the frequency of the $L_Z$--$V_R$  wave from \citet{Antoja2022}. As in our original work, A18, these times are consistent with a previous pericentre passage of Sagittarius.
    \item The mentioned trends with angular momentum and height are robust to using frequencies for hotter orbits, different potential models, different distance determinations, and possible (small) distance biases. 
\end{itemize}

  We note that our modelling of the phase spiral could be oversimplistic. For example, we only consider the crossings of the phase spiral with the $Z=0$ and $V_Z=0$ axis but it would be better to use the continuity of the entire phase spiral. This would be more in line with the work of \citet{Widmark2021}. Recently, we found out that studies carried out simultaneously to ours \citep{Frankel2023,DarraghFord2023} used action-angle variables in an assumed potential to unwind the spiral into a `straight' line for the time fit. While our method does not use the full information of the phase spiral at the same time, it confers a slight advantage in that it does not require the assumption of a potential until the final steps, making it easier to use in a combined fit of the phase-mixing time and potential. In any case, these studies also find different time determinations at different angular momenta (between 0.2 and 0.6 Gyr in \citealt{Frankel2023}; 0.3 and 1 Gyr \citealt{DarraghFord2023}) and with different radius and azimuth \citep[between 0.3 and 0.8 Gyr in][]{Widmark2022a}, which is similar to our results.

These differences in the obtained times for different angular momenta and heights are not expected in the simplest interpretation of the phase spiral coming from a phase-mixing process. Below we discuss different explanations. 

It could be that we are seeing differences in the time of response to the perturbation of different disc regions. However, for the case of the Sagittarius impact studied by \citet{Gandhi2022}, the authors found differences in the onset of the perturbation effects at different radii of the order of 100 Myr, which is smaller than what we measure. In addition, they do not see a clear trend with guiding radius.  \citet{Widmark2022b} tested their machinery of determining the potential from the phase spiral shape on the N-Body simulation from \citet{Hunt2021} and found time differences in small ranges of $R$ of the order of 100-200 Myr. In test-particle simulations, \citet{DarraghFord2023} find slightly larger ($\approx 200$ Myr) time variations with position (guiding centre coordinates). In addition, the time differences with $Z$ at a fixed angular momentum would be hard to explain with this hypothesis alone, although as discussed above, our method could be affected by bias for large $Z$. The spatial and temporal propagation of different bending waves remains poorly studied and we plan to examine this with various models in the future.

Another explanation is that there is a mismatch between the vertical and radial dependence of the potential in the models (that we used to compute the vertical frequencies) and in the real MW. We also find possible evidence for this mismatch by considering and comparing the coordinates of the phase spiral when crossing the $Z=0$ and the $V_Z$ axis (\app\ref{app_test}).  One could infer the potential that could make these trends disappear. For example, the linear trend seen with $L_Z$ (dashed line in \f\ref{f_T}) could be used to `correct' the potential (the gradient in frequencies) in order to obtain similar times. This is not trivial because of the degeneracy and the possible self-gravity effects. 
 
Indeed, the time variations could also be evidence for self-gravity acting differently in different parts of the disc. Self-gravity tends to amplify the phase-mixing times \citep{Darling2019} and in particular could be of less importance in the outer parts of the disc \citep[e.g.][]{Shen2006}. This could mean that the times that we obtain in the outer parts, that is the larger times, are closer to the true perturbation time. However, the radial behaviour of different kinds of bending waves that originate from different perturbations might be different (e.g. bending waves associated to the bar or due to external torques from galactic satellites or even from a misaligned halo) and this possibility has not been thoroughly studied. Recently, \citet{DarraghFord2023} found large offsets between the interaction time and the recovered time in their N-Body simulation using their action-angle modelling, which they attribute to self-gravity or the effects of the associated halo wake.
 
  Finally, our finding could be related to other aspects of the disc dynamics and how they interact with each other. By this, we mean the effects of the bar, the spiral arms, the warp, and different bending and breathing modes, which could have acted or be acting in addition to the vertical perturbation causing the phase spiral. For instance, \citet{Widmark2022d} recently showed that some features of the vertical density and velocity seem related to the local spiral arm, while theoretical studies have also led to the identification of vertical velocity effects in spiral arm models \citep[e.g.][]{Faure2014,Kumar2022}. In addition, if the disc mid-plane oscillates because of one of these additional distinct phenomena, the $Z$--$V_Z$ centroid of the phase spiral will oscillate and affect the time determinations (Appendix~\ref{app_test}).

In conclusion, we are now able to see a higher level of complexity in the morphology of the vertical projection, as well as unexpected trends, possibly indicating different perturbations (perhaps from different agents or different times), as already pointed out by \citet{Hunt2022} ---following the discovery of the double phase spirals---, and/or complex phase mixing processes. 
The variations that we see in our derived phase-mixing times with vertical position and angular momentum likely indicate inadequate modelling due to uncertainties in the potential model for the Galaxy; our neglect of self-gravity; the existence and interaction of multiple perturbations, or a combination of these. Although the new \Gaia DR3 data definitively bring us a clearer picture of the MW phase spirals, there is still much to be understood and modelled.

%

\begin{acknowledgements}
    Project supported by a 2021 Leonardo Grant for Researchers and Cultural Creators, BBVA Foundation. The BBVA Foundation accepts no responsibility for the opinions, statements and contents included in the project and/or the results thereof, which are entirely the responsibility of the authors. This work was (partially) supported by the Spanish MICIN/AEI/10.13039/501100011033 and by "ERDF A way of making Europe" by the “European Union” through grant PID2021-125451NA-I00, and the Institute of Cosmos Sciences University of Barcelona (ICCUB, Unidad de Excelencia ’Mar\'{\i}a de Maeztu’) through grant CEX2019-000918-M.
      TA acknowledges the grant RYC2018-025968-I funded by MCIN/AEI/10.13039/501100011033 and by ``ESF Investing in your future''. 
       BGC acknowledges  financial support from the Spanish Ministry of Economy and Competitiveness (MINECO) under grant number RTI2018-096188-B-I00 and IPARCOS Institute for the grant "Ayudas de doctorado IPARCOS-UCM/2022". MB received funding from the University of Barcelona’s official doctoral program for the development of a R+D+i project under the PREDOCS-UB grant. CL acknowledges funding from the European Research Council (ERC) under the European Union’s Horizon 2020 research and innovation programme (grant agreement No. 852839). 
      This work has made use of data from the European Space Agency (ESA) mission
{\it Gaia} (\url{https://www.cosmos.esa.int/gaia}), processed by the {\it Gaia}
Data Processing and Analysis Consortium (DPAC,
\url{https://www.cosmos.esa.int/web/gaia/dpac/consortium}). Funding for the DPAC
has been provided by national institutions, in particular the institutions
participating in the {\it Gaia} Multilateral Agreement.

\end{acknowledgements}

%
   \bibliographystyle{aa} 
   \bibliography{mybib} 
%
\begin{appendix}\label{app}


\section{Queries used to select samples}
\label{app_queries}

In this Appendix, we show a few examples of queries to the \Gaia Archive \url{https://gea.esac.esa.int/archive/} to retrieve the data:

\begin{lstlisting}[title=A,label={q1},caption=This query is used to  retrieve stars with radial velocity with additional selection in temperature and $\tt{RUWE}$ (see main text) for our basic sample.]
SELECT gaia.source_id,gaia.ra,gaia.dec,
gaia.parallax,gaia.pmra,gaia.pmdec,
gaia.radial_velocity,gaia.parallax_error,
gaia.phot_g_mean_mag,gaia.bp_rp,
gaia.grvs_mag,gaia.rv_template_teff
FROM gaiadr3.gaia_source AS gaia 
WHERE gaia.radial_velocity is not null   AND gaia.ruwe < 1.4 AND gaia.rv_template_teff<8500
\end{lstlisting}

\begin{lstlisting}[title=A,label={q2},caption=This query is used to retrieve the $\tt{fidelity}$ flag (see main text) that is later merged by $\tt{source\_id}$ with the data from query \ref{q1} for additional selection.]
SELECT Rybizki.source_id, Rybizki.fidelity_v2
FROM external.gaiaedr3_spurious AS Rybizki
JOIN gaiadr3.gaia_source AS gaia  ON gaia.source_id = Rybizki.source_id
WHERE gaia.radial_velocity is not null  
\end{lstlisting}

\begin{lstlisting}[title=A,label={q3},caption=This query is used to retrieve the BJ distance (see main text) that is later merged by $\tt{source\_id}$ with the data from query \ref{q1}.]
SELECT CBJ.source_id, CBJ.r_med_photogeo
FROM external.gaiaedr3_distance AS CBJ 
JOIN gaiadr3.gaia_source AS gaia  ON gaia.source_id = CBJ.source_id
WHERE gaia.radial_velocity is not null 
\end{lstlisting}

\begin{lstlisting}[title=B,label={q4},caption=This query is used to retrieve the $\tt{source\_id}$ of giants stars that is later merged by $\tt{source\_id}$ with the data from query \ref{q1}.]
SELECT g.source_id
FROM gaiadr3.gaia_source AS g
WHERE (g.teff_gspphot<5500 and g.teff_gspphot>3000)
and (g.logg_gspphot<3.) and (g.radial_velocity is not null)
\end{lstlisting}

\section{Different distances}\label{app_dist}

\begin{figure}
   \centering
   \includegraphics[width=1.\columnwidth]{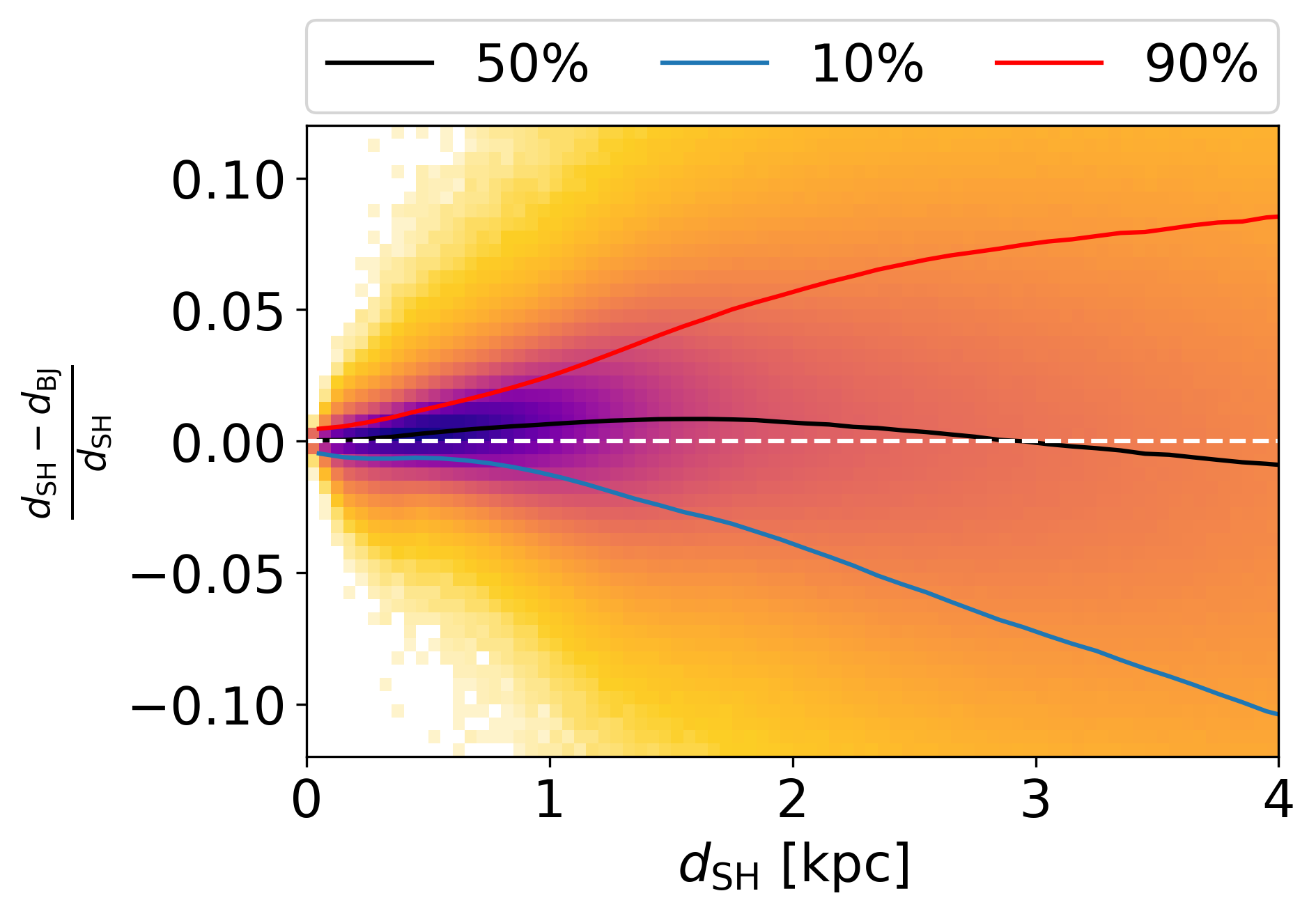}
   \caption{Comparison between sets of distances. Relative differences between distances from SH (\citealt{Anders2022}) and BJ distances  (\citealt{BailerJones2021}) as a function of SH distances for the  stars in our sample for which these two distances are available. The black line shows the median of the distribution, while the blue and red lower and upper lines show the 10 and 90 percentiles (i.e. they enclose 80\% of stars).}
    \label{f_dis}%
\end{figure}

\begin{figure}
   \centering
   \includegraphics[width=1.\columnwidth]{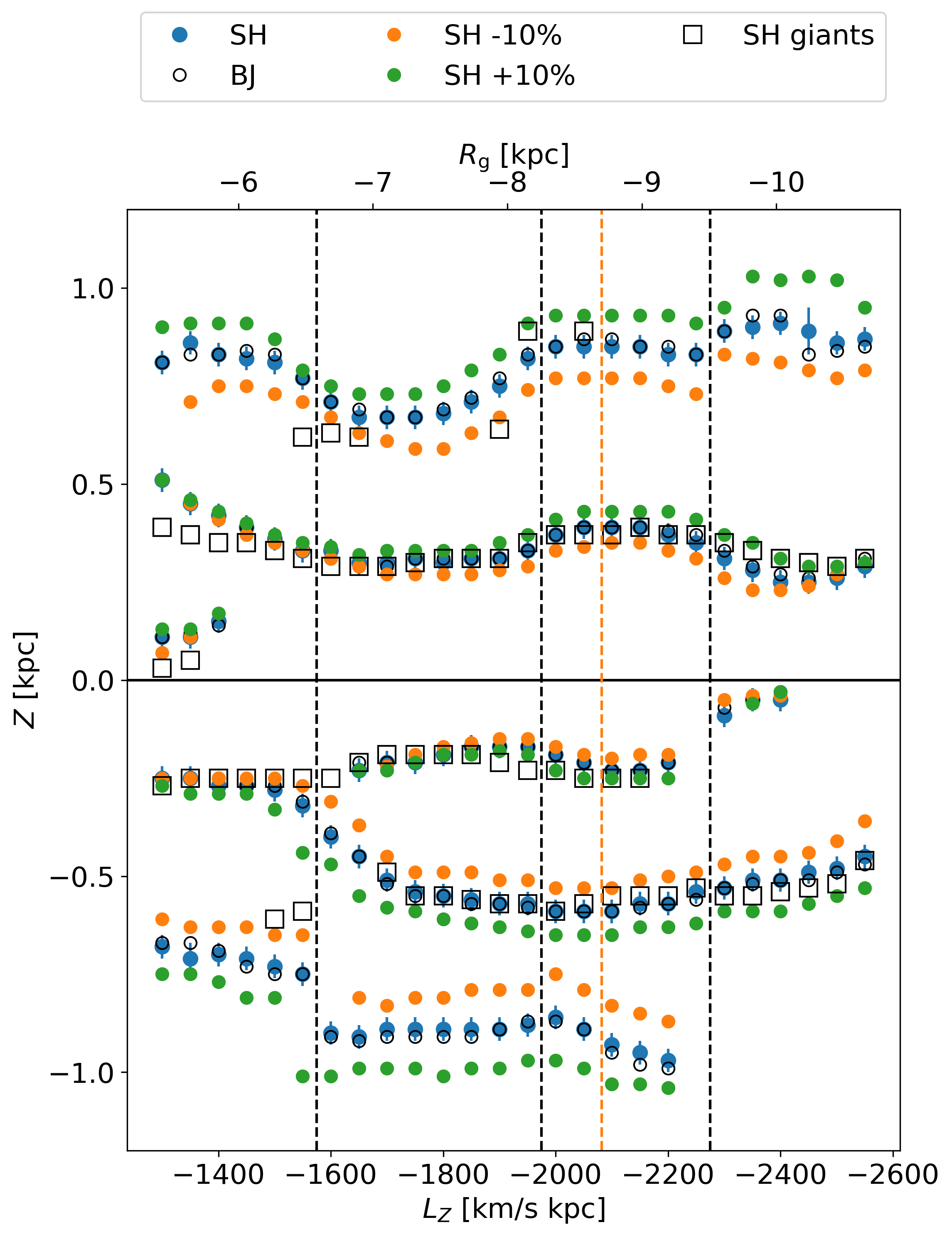}
   \caption{Crossing positions of the phase spiral for different distances. The blue circles and error bars mark the $Z$ coordinate for $V_Z=0$ (\zc) of the phase spiral determined from the edge detector for the SH sample. The other symbols indicate the crossings for the BJ distances, for distance biases of $\pm10\%$, and for a selection of giant stars.}
    \label{f_Zc2}%
\end{figure}

Here we compare the SH and the BJ distances of the 25\,385\,209 stars that remain after the cuts specified in \s\ref{s_data}. This is shown in \f\ref{f_dis}. On average, there is a negligible bias between distances. For example, the relative differences are of 0.8\% at around 1.5 kpc and of -0.9\% at 4 kpc, that is smaller than 1\% for the distances we are dealing with. For 80\% of the stars, the differences are smaller than 11\% up to 4 kpc. Indeed, when we calculate our \zc{} using the different distance sets, our values change only very slightly (\f\ref{f_Zc2}) and, consequently, our derived times do not change significantly either (\f\ref{f_Td}).

\section{Effect of distance bias on the parameters of the phase spiral}
\label{appb_biases}

Given that $Z$ and $V_z$ can be written as

\begin{equation}\label{eq_z}
    Z = d\,sin(b)\, \mathrm{and}
\end{equation}

\begin{equation}\label{eq_Vz}
    V_z = \kappa\,d\,\mu_b\,cos(b) + V_{los}\,sin(b),
\end{equation}

\noindent we can make some approximate calculations to understand how the distances affect the shape of the phase space spiral.

Assuming that the effect of the distance bias only manifests at a certain distance, and that the range in Galactic latitude for disc stars shrinks with this distance, we can neglect the contribution of the line-of-sight velocity to $V_z$ without loss of generality. We can further assume that the errors in the sky position are negligible, and that proper-motion errors are perfectly Gaussian. If we do so, it is trivial to prove that, on average, 

\begin{equation}\label{eq_z_bias}
    Z \simeq Z^{true}\frac{d}{d^{true}}    
,\end{equation} 

\noindent and, similarly, 

\begin{equation}\label{eq_Vz_bias}
    V_z \simeq V_z^{true}\frac{d}{d^{true}}.   
\end{equation}

This result implies that both coordinates will suffer from a bias in the distance in a similar manner, at any position in the disc. For instance, if the distances are underestimated by 10\%, the phase space spiral will shrink by the same amount, thus keeping the proportions. However, we note that even if proportions are kept, the pitch angle will change and become larger if the spiral shrinks, and smaller if the spiral grows.

In general, we know that the bias increases with distance for most distance estimators based fully, or even partially, on the parallax. Therefore, the amount by which the spiral shrinks or grows will change as a function of $Z$ for a certain $X$ and $Y$, as the heliocentric distance will increase with $Z$. Therefore, the deformation will not be as simple as a scaling of the original one.

Moreover, the process of estimating a distance from a parallax for one star usually leads to a probability distribution of distances that is not symmetric. Therefore, rather than having a simple bias, normally the mode of the resulting distance probability distribution is smaller than the true distance (which should coincide with the median), causing in turn a long tail towards large distances \citep{Luri2018}. This further complicates any attempt to predict the appearance of the phase spiral. 

To test these deductions, first we generate particles along a perfect Archimedean spiral in a manner such that the number of particles increases with the distance from the centre, which we do simply for visualisation purposes\footnote{The area covered by a constant angular step increases along the spiral, as the distance from the centre grows, which causes the consequent drop in density we try to counteract by drawing the angles of the particles along the spiral from a Beta probability distribution function with a=2 and b=5. In other words, the angles along the spiral are not drawn uniformly.}. Once we have this \emph{true} spiral, we place it at a certain heliocentric distance of our choice (this is the distance of the particles at Z=0), and then calculate the true distance of all the particles in it by inverting Eqs. \ref{eq_z} and \ref{eq_Vz}. We then draw a parallax for each particle from a normal distribution centred at its true parallax (inverse of its true distance), with an error set to a certain parallax over error.

\begin{figure}
   \centering
   \includegraphics[width=0.7\linewidth]{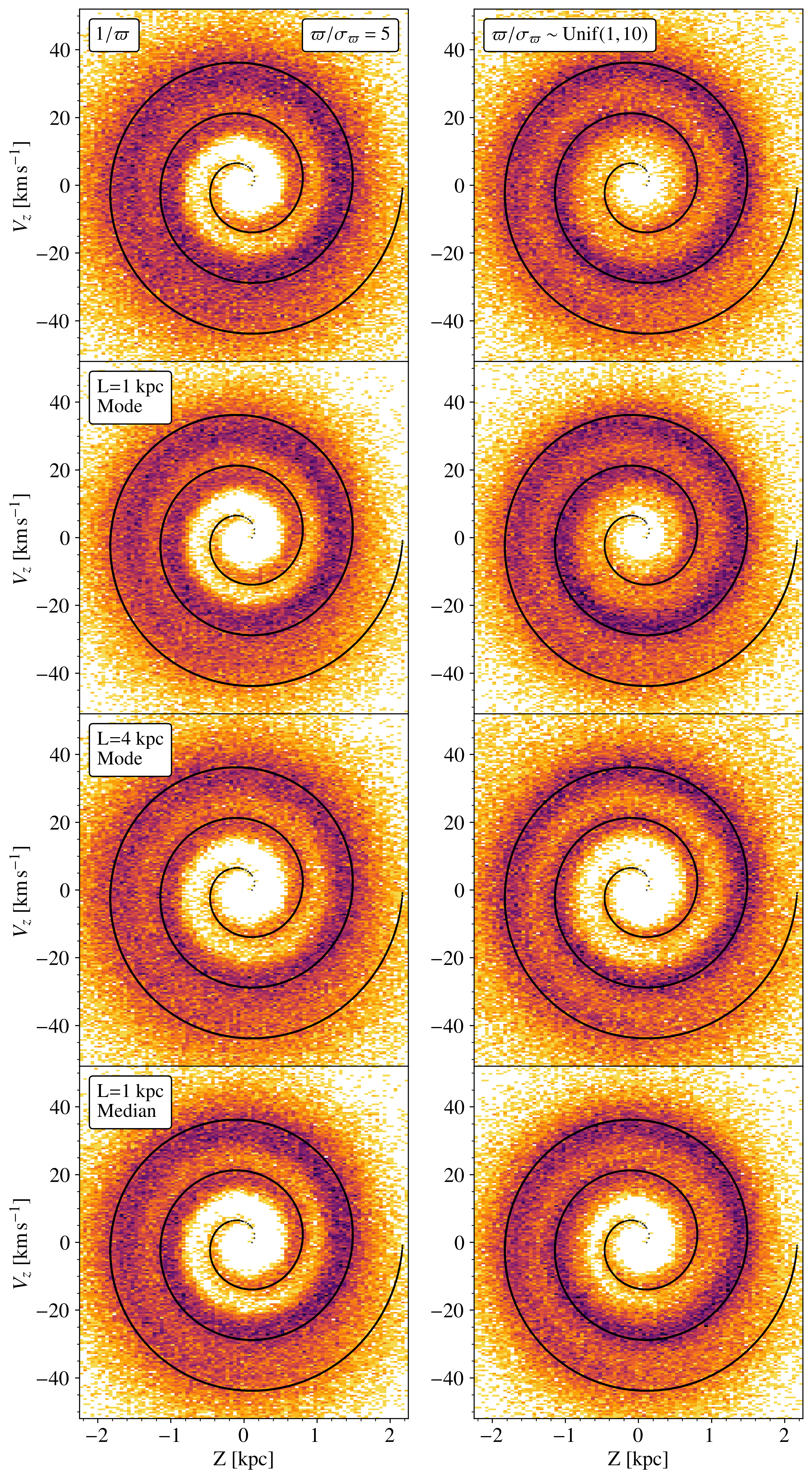}
   \caption{Effects of the distance bias on a perfect Archimedean spiral. For all panels, the centre of the spiral is at 4$\,\kpc$ from the Sun, and the black curve is the true spiral. Left column: Mock particles sampled with a constant parallax over error of 5. Right column: Same but the parallax over error is sampled uniformly from 1 to 10. Rows correspond to different distance estimates, from top to bottom: parallax inversion; Bayesian estimate with L=1$\,\kpc$ and taking the mode of the posterior; same as above but with L=4$\,\kpc$; Bayesian estimate with L=1$\,\kpc$ but taking the median of the posterior.}
    \label{f_fakespiral_poe5}%
\end{figure}

The left column of Fig. \ref{f_fakespiral_poe5} shows the effects of assuming a fixed parallax over error of 5 for all particles on a spiral that is located at 4$\,\kpc$ from the Sun. The different rows correspond to different distances estimates: inverting the parallax or using the \citet{Bailer-Jones2015} exponentially decreasing volume prior with different parameters. As we can see, the peak of the density is displaced inwards as expected (see above), with a long tail outwards accompanying it that blurs the signal. The right column of Fig. \ref{f_fakespiral_poe5} is similar to the left column, but now we have sampled the parallax over error randomly in the range from 1 to 10. This results in a less biased and slightly more blurred  spiral. The differences between rows (distance estimates) here is almost unnoticeable, although using the median (bottom-right panel) instead of the mode of the posterior (second and third rows on the right) seems to provide less biased results. However, in general,  we will have a mixture of populations that depends on the location with respect to the Sun of the bin we are sampling, that is on the magnitude and colour distribution of the stars in that subsample and their heliocentric distances. Therefore, even if we cannot quantify the exact deformation of the spiral, it seems reasonable to expect it to shrink.

As a second test, we now recompute the phase-space coordinates of the full sample after enlarging and decreasing their distances by 10\%. These two new dummy samples help us, to begin with, to confirm that the approximations presented in Eqs. \ref{eq_z_bias} and \ref{eq_Vz_bias} are valid in general. More importantly, with the new sample, we can redo all calculations of our analysis. First in \f\ref{f_Zc2}, we show the \zc{} of the phase spiral when distances are decreased by 10\% (orange circles) and increased by 10\% (green) which can be compared to the unscaled SH distances (blue circles). In general, we find that the bias mostly reproduces our expectations, that is  the \zc{} are scaled by $\pm10\%$. Some points differ from these expectations, which might be due to selection effects.  \f\ref{f_Td} shows the recomputed phase mixing times for these new \zc{} and we find that, in most cases, they fall within the errors of the fiducial case (in 80\% of the cases). More details are given in the main text. We note that although the biased \zc{} do not fall within the errors of the initial (unbiased) \zc{}, this is not the case for the time determinations. This is because, to compute a time and its error bar, we use two consecutive \zc{} and combine the different upper and lower limits of these two \zc{} to obtain the minimum and maximum frequency differences that give rise to the maximum and minimum times, respectively.

\section{Selection effects}
\label{app_sf}

In this Appendix, we examine the possible effects of the selection function of our samples in the determined \zc{} and phase-mixing times. The \Gaia selection function is definitively a complicated matter. A possible worry for our work is that the
characteristics of our sample, which certainly change with $R$ and $Z$, bias our results. 
At different parts of the phase spiral and at different angular momenta, the dominating population might change. For example, the proportion of thick-disc over thin-disc stars or the average age can vary with radius. In particular, the average radial action of stars might change. We used actions computed in \citet{RecioBlanco2022} to check that mean values of the radial action in the $L_Z$ and $Z$ ranges explored here go from about 20 to 80 $\kms\kpc$ (the $75$ percentile can reach 120 $\kms\kpc$). However, here we find that changing from $J_R=0$ to $200\,\kms\kpc$ gives time differences mostly in the range of $[0.04, 0.16]$ (Table~\ref{t_T}), which is smaller than the systematic differences seen with angular momentum and height (\f\ref{f_T}). The trends of the derived times with angular momentum (and height) are hard to explain with only selection effects.

\begin{figure}
\centering
   \includegraphics[width=1.\columnwidth]{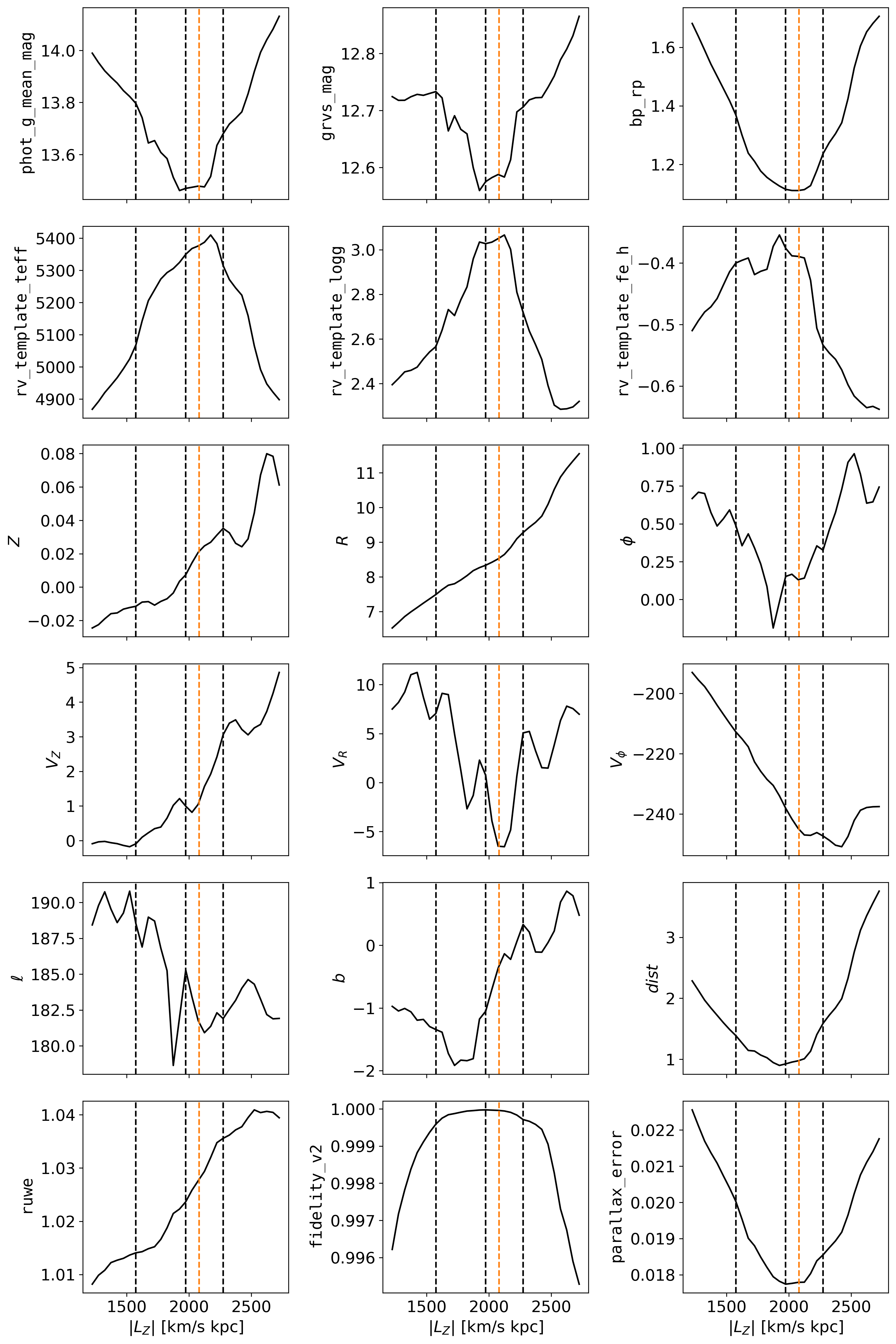}
   \caption{Investigation of some selection function effects. Mean values of different variables as a function of angular momentum $L_Z$. The dashed vertical lines indicate some positions where we see jumps in the shape of the phase spiral and/or the determined phase mixing times. The orange dashed line is the angular momentum of the Sun.}
    \label{f_sf}%
\end{figure}

We nevertheless explore the mean value of different variables as a function of $L_Z$ for the sample of stars with $|\phi|<20\deg$ that is used in most of our analysis. This is shown in \ff\ref{f_sf}. Most of the quantities show a similar pattern (peak or valley at the the Sun's angular momentum, orange vertical dashed line) that is explained by having increasing distance for values of the $|L_Z|$ departing from that of the Sun (right panel in the fifth row). It is not straightforward to translate this trend into a trend in $L_Z$ (or in $Z$) because it is the difference between frequencies that enters in the calculation, but the peak and valley trends make the  time increase with $|L_Z|$ hard to explain. By contrast, the hypotheses given in the discussion in the main text (incorrectness of the potential models, neglect of self-gravity, etc.) are easier to explain and are somewhat expected causes.

In all panels of \ff\ref{f_sf}, we use vertical black dashed lines to mark some positions of discontinuity in the \zc{} ($L_Z=-1575$ and -2275 $\kms\kpc$, also marked in \f\ref{f_Zc2}) and in the time determinations ($L_Z=-1975\,\kms\kpc$). Some of the lines could be correlated with jumps for example in the average magnitude of stars (two first panels of the first row, but with very small difference in magnitude) or with slight changes in the trends of colour (right panel in the first row). We also see some correlation with the known wave in $VR$ (middle panel in the fourth row). At this point, we cannot therefore discard that the jumps are selection effects or conclude that they are real physical effects.

\section{Additional tests}
\label{app_test}

\begin{figure}
\centering
   \includegraphics[width=1.\columnwidth]{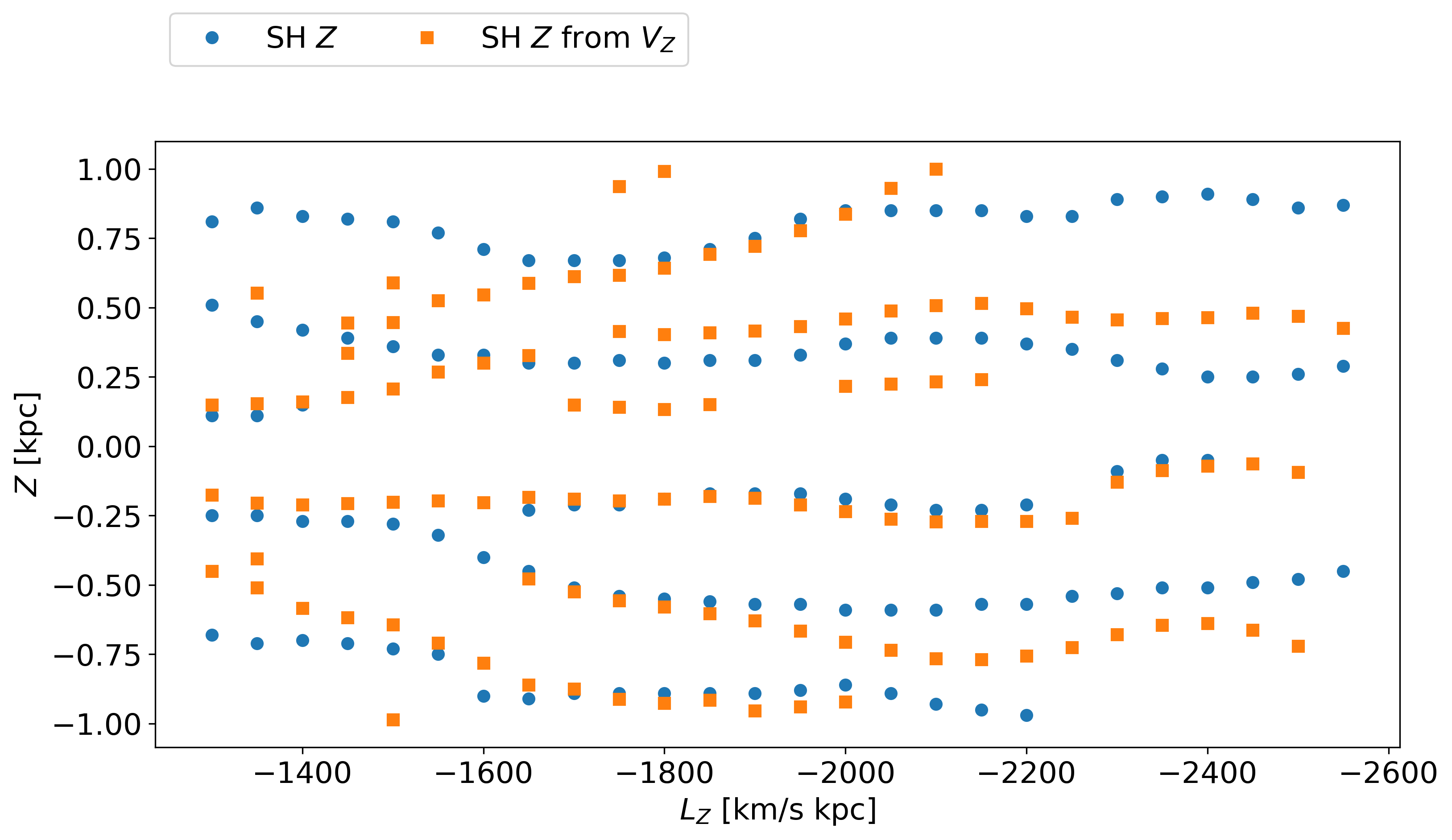}
   \caption{Crossing positions of the phase spiral in the $V_Z=0$ and $Z=0$ axis. The blue circles mark the $Z$ coordinate of the phase spiral at $V_Z=0$ (\zc) determined from the edge detector for the SH sample. The orange squares indicate the \zc{} corresponding to the crossings \vzc{} at the $Z=0$ axis (see text).}
    \label{f_VZ}%
\end{figure}

 Here, we examine the consecutive $V_Z$ crossings of the phase spiral at $Z=0$, which we name \vzc. To determine the \vzc{} we use the same method as for \zc{} based on the edge detector. We then find the maximum $Z$ corresponding to an orbit with that maximum $V_Z$ under the fiducial potential. We do this using the same orbital integrations carried out to determine the vertical frequency, for which, at each $L_Z$, we have a pair of maximum $Z$ and $V_Z$. This can be thought of as obtaining a new set of \zc{} that can be compared to the previous ones. We can examine whether these new \zc{} sequences run through the middle of the previous \zc{} sequences, as expected from the most simple interpretation of the phase spiral. The results of this test (\f\ref{f_VZ}) show that this is not fulfilled when we assume our fiducial potential, as we see some sequences overlapping. Our understanding is that this might have several implications, with the simplest one being that our assumed potential is incorrect. Other interpretations could be an underlying complex bending wave, time dependence of the global potential, or the existence of multiple interfering phase spirals.

\begin{figure}
\centering
   \includegraphics[width=1.\columnwidth]{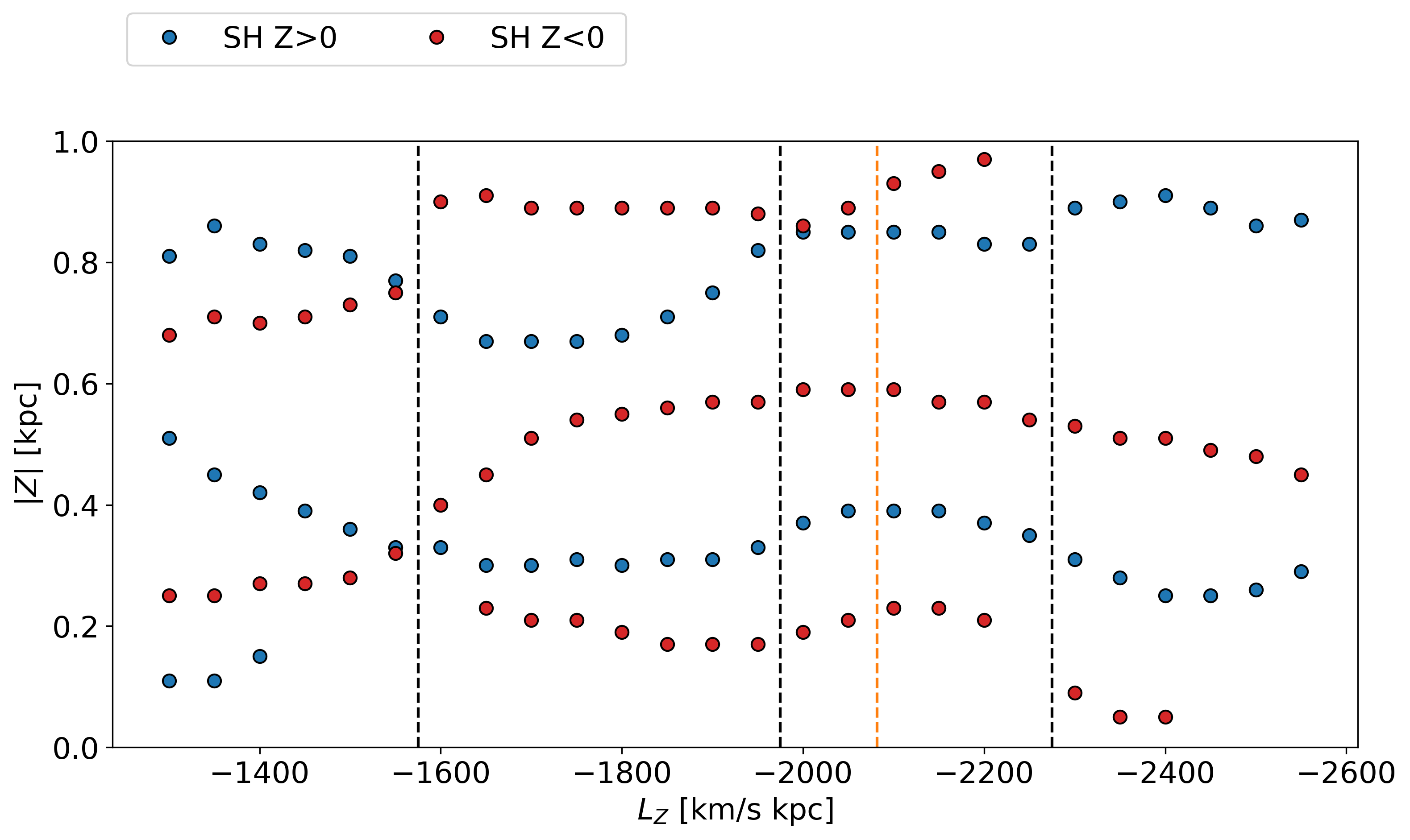}
   \caption{Crossing positions of the phase spiral for $Z>0$ and $Z<0$. The blue and red circles mark the $|Z|$ coordinate of the phase spiral at $V_Z=0$ (\zc) for positive and negative $Z$, respectively.}
    \label{f_Zposneg}%
\end{figure}

We run a second test in which we compare the \zc{} for either $Z>0$ or $Z<0$. Again in the most simple interpretation of the phase spiral, we should expect that $|\zcm|$ sequences at $Z<0$ (i.e. when we consider the absolute value of \zc{} from crossings of the phase spiral at $Z<0$) run approximately through the middle of the \zc{} sequences at $Z>0$. This is studied in \f\ref{f_Zposneg} and we see that the red sequences (from $Z<0$) are roughly but not exactly in the middle of the blue sequences (from $Z>0$). For instance, there are  crossings of red and blue sequences at certain positions, where we already identified jumps or undulations. This could be evidence of the existence of a bending wave in the disc that is superimposed over the phase spiral patterns. However, if this is the case, it does not appear as a simple wave that can be easily subtracted. Also, in the case of having a bending wave, the reference system, the potential, and the vertical frequencies would be somehow ill-defined concepts, and thus our modelling would not be appropriate. On the other hand, this could be evidence of more complex phase mixing, as the rest of our study also suggests.

\section{Additional material}
\label{app_rc}

\begin{figure}
   \centering
   \includegraphics[width=1.\columnwidth]{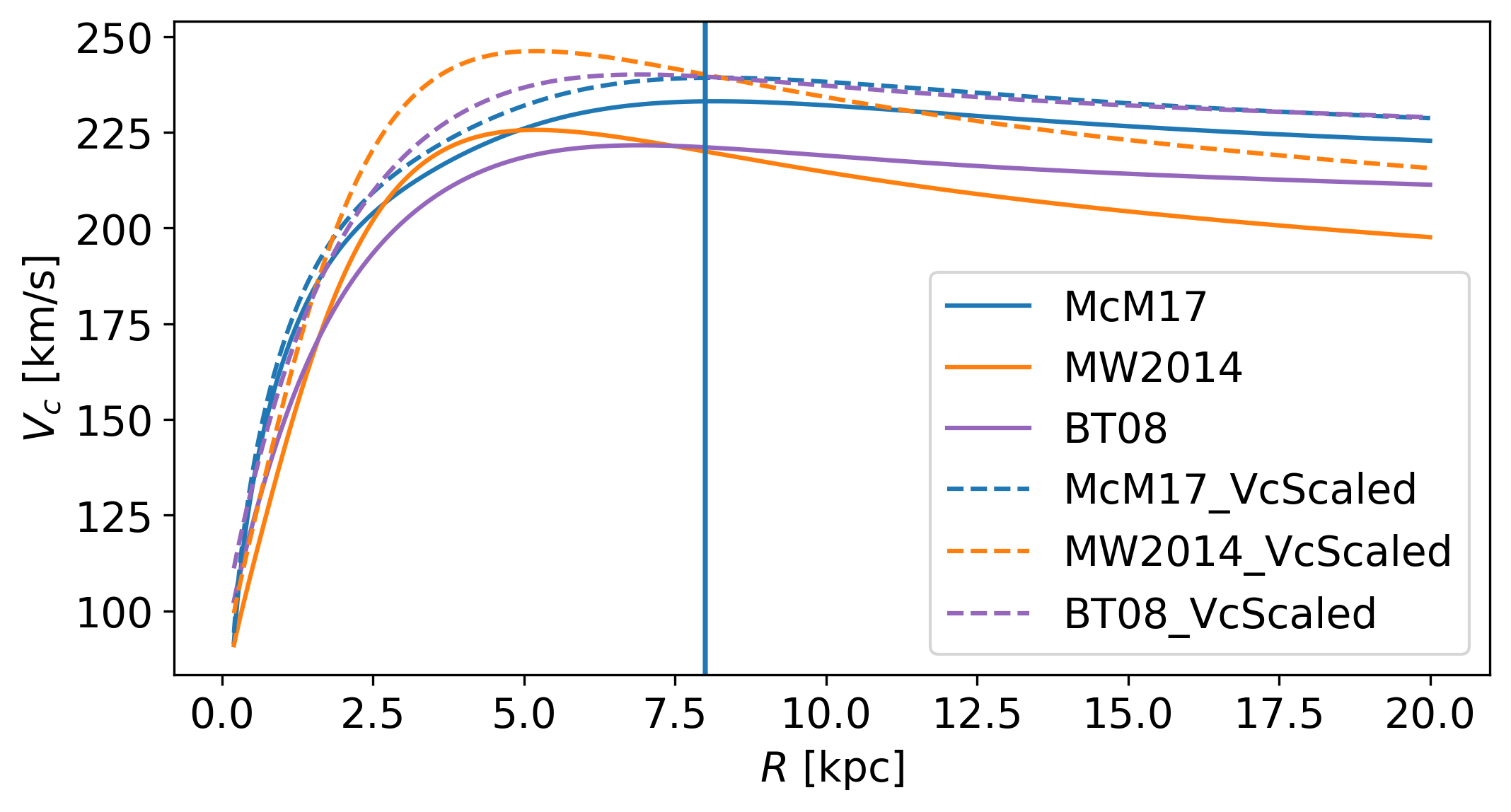}
   \caption{Circular velocity curves of the different potentials used.}
    \label{f_rc}%
\end{figure}

\begin{figure}
   \centering
   \includegraphics[width=1.\columnwidth]{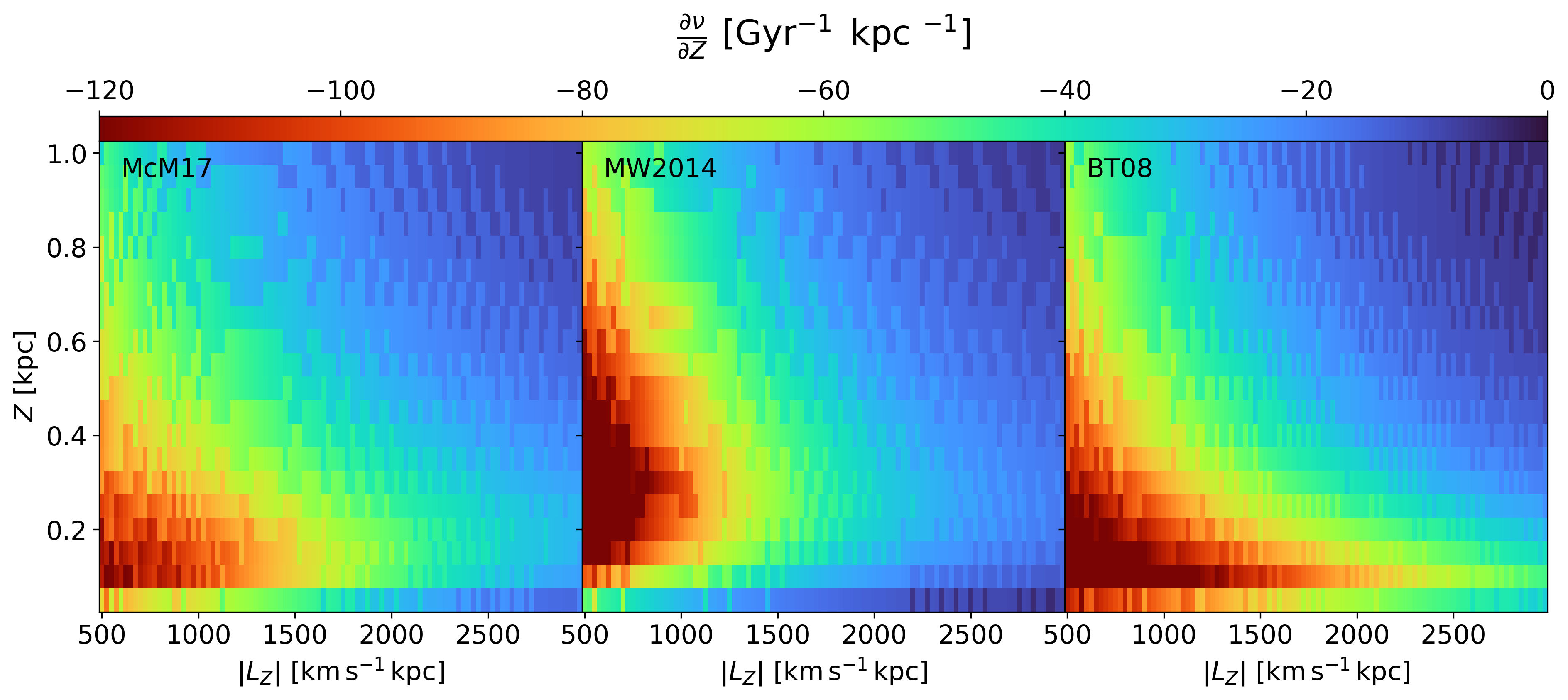}
   \caption{ Vertical gradient of the vertical frequencies for the different potentials used.}
    \label{f_fg}%
\end{figure}

Here we present the different gravitational potential models for the MW used in our study (\f\ref{f_rc}) and their respective vertical gradients in the vertical frequencies (\f\ref{f_fg}).

\end{appendix}

\end{document}